\documentclass[aps,pre,reprint]{revtex4-1}

\usepackage{graphicx,tabularx}
\usepackage{epsfig}
\usepackage{amsmath,amsfonts,amssymb,amsthm,bbm}
\usepackage{url}
\usepackage{hyperref}
\usepackage{rotating}

\theoremstyle{plain}
\newtheorem{thm}{Theorem}
\newtheorem*{thm*}{Theorem}

\newtheorem*{lem*}{Lemma}
\newtheorem{cor}{Corollary}
\newtheorem*{cor*}{Corollary}
\newtheorem{prop}{Proposition}
\newtheorem*{prop*}{Proposition}

\theoremstyle{definition}

\newtheorem{Def}{Definition}

\def\cR{{\mathcal{R}}}
\def\cN{{\mathcal{N}}}

\def\cT{{\mathcal{T}}}

\def\cL{{\mathcal{L}}}

\def\be{\begin{equation}}
\def\ee{\end{equation}}
\def\ben{\begin{equation*}}
\def\een{\end{equation*}}

\begin{document}

\title{Critical Tokunaga model for river networks}

\author{Yevgeniy Kovchegov}
\email[]{kovchegy@math.oregonstate.edu}
\thanks{This work is supported by NSF award DMS-1412557.}
\affiliation{Department of Mathematics, Oregon State University, Corvallis, OR, 97331-4605, USA}

\author{Ilya Zaliapin}
\email[]{zal@unr.edu}
\thanks{This work is supported by NSF award EAR-1723033.}
\affiliation{Department of Mathematics and Statistics, University of Nevada, Reno, NV, 89557-0084,
USA}

\author{Efi Foufoula-Georgiou}
\email[]{efi@uci.edu}
\thanks{This work is supported by NSF awards DMS-1839336 and EAR-1811909.}
\affiliation{Departments of Civil and Environmental Engineering and Earth System Science,\\ University of California, Irvine, CA 92697.}

\date{\today}

\begin{abstract}
The hierarchical organization and self-similarity in river basins have been topics of extensive research in hydrology and geomorphology starting with the pioneering work of Horton in 1945. Despite significant theoretical and applied advances however, the mathematical origin of and relation among Horton laws for different stream attributes remain unsettled.  
Here we capitalize on a recently developed theory of random self-similar trees to introduce a one-parametric family of self-similar critical Tokunaga trees that elucidates the origin of Horton laws, Hack's laws, basin fractal dimension, power-law distributions of link attributes, and power-law relations between distinct attributes. 
The proposed family includes the celebrated Shreve's random topology model and extends to trees that approximate the observed river networks with realistic exponents. 
The results offer tools to increase our understanding of landscape organization under different hydroclimatic forcings, and to extend scaling relationships useful for hydrologic prediction to resolutions higher that those observed. 
\end{abstract}

\pacs{}
%\keywords{}

\maketitle

%====================================================
\section{Introduction}\label{sec_intro}
In a pioneering study ``{\it of streams and their drainage basins}'', Robert E. Horton \cite{Horton45} introduced the concept of river stream order and formulated two fundamental laws of the composition of stream-drainage nets. The {\it Law of Stream Numbers} postulates a geometric decay of the numbers $N_K$ of streams of increasing order $K$, with the exponent $R_B$. The {\it Law of Stream Lengths} postulates a geometric growth of the average length $L_K$ of streams of increasing order $K$, with the exponent $R_L$. During the 20th century, geometric dependence on the stream order (conventionally referred to as {\it Horton law}) has been documented for multiple stream attributes, including upstream area, magnitude (number of upstream headwater channels, also called sources), the total channel length, the longest stream length, link slope, mean annual discharge, energy expenditure, etc. \cite{RIR01}. Despite their elemental role in describing the key regularities in river stream networks (such as fractal dimension, Hack's law, etc.), Horton's laws remain an empirical finding and their origin and apparent ubiquity remain unsettled \cite{Kirchner93}.

The first attempt at a rigorous explanation of Horton laws was made by Ronald L. Shreve \cite{Shreve66} in the 1960s, who examined a ``{\it topologically random population of channel networks}'', where all topologically distinct networks with given number of first-order streams are equally likely. This model is equivalent to the celebrated critical binary Galton-Watson branching process with a given population size \cite{BWW00,Pitman}. Shreve's calculations imply that in this model the Horton law of stream numbers holds with $R_B = 4$. Although not attempted by Shreve, it can be shown \cite{BWW00} that the law of stream lengths also holds here with $R_L = 2$ under the assumption of constant or identically distributed edge lengths. Albeit insightful and mathematically tractable, the random topology model deviates from observations, which became apparent with the development of improved methods for extracting river networks \cite{DBE94,Pec95}. 
This called for developing alternative modeling approaches for river networks. 

It has proven challenging to find a model that would be mathematically tractable and flexible enough to reproduce the Horton exponents and other scaling laws observed in river basins. One end of the modeling spectrum is occupied by conceptual approaches, such as Peano fractal basin (\cite{RIR01}, their Sect. 2.4; \cite{Horton45}, their Fig. 25) or Scheidegger's lattice model \cite{Sch67,TNT88}. These models provide an invaluable insight into the origin of the observed scalings; they however lack realistic dendritic patterns and values of scaling exponents. On the other end are simulation approaches that can generate visually appealing networks that closely fit selected exponents, but can be analytically opaque. The Optimal Channel Network (OCN) model \cite{RRR+92,RRR+93,RRRIB93,MRR+96,RRBMR14,BBB+18} is a well-recognized simulation technique. Following the energy expenditure minimization principle, the model creates random drainage basins on a planar lattice (or more general graphs). We refer to \cite{RIR01} for a comprehensive discussion of these and other models. 

Despite the progress achieved by the modeling efforts of the $20^{th}$ century, the following essential questions remain unanswered: 
{\it What are sufficient conditions for Horton laws? 
What are the values of the Horton exponents for river basins? 
How are the Horton exponents for different stream attributes related to each other and to other basin parameters?} 
There is a consensus that Horton's laws are connected to the self-similar structure of a basin \cite{RIR01,Pec95,GW89,GTD07}, which is generally understood as invariance of basin's statistical structure under changing the scale of analysis (zooming in or out) \cite{Turcotte_book}. However, a consensus is still lacking about a suitable mathematical definition of tree self-similarity. Three alternative definitions have been proposed: the Toeplitz property of the Tokunaga coefficients \cite{Pec95, NTG97}; the invariance of a tree distribution with respect to the Horton pruning (cutting source streams) \cite{BWW00}; and statistical self-similarity of basin attributes \cite{GW89,PG99}. The related unsettled questions are: {\it How are the alternative definitions related?} and {\it Is self-similarity (any version) sufficient for selected Horton laws?} These questions extend to other areas beyond hydrogeomorphology where Horton laws and related scalings have been reported, including computer science \cite{F+79,DP06}, statistical seismology \cite{BP04,HTR08,Y13,ZBZ13}, vascular analysis \cite{Kassab00}, brain studies \cite{Cetal06}, ecology \cite{CLF07}, and biology \cite{TPN98}. 

We answer the above questions within a self-consistent mathematical theory of random self-similar trees recently developed by the authors \cite{KZ20survey}. 
The theory develops the concept of tree self-similarity that unifies the alternative existing definitions, rigorously explains the appearance and parameterization of Horton laws, and offers a novel approach to modeling a variety of dendritic systems. The goal of this paper is to adapt and extend the theory to the studies of river networks. Most notably, we show that two fundamental and practically appealing properties -- {\it coordination} and {\it Horton prune invariance} -- result in trees that respect a wealth of scaling laws observed in landscape dissection. Furthermore, we propose a new one-parametric family of {\it critical Tokunaga trees}, which reproduces multiple Horton laws and related scalings reported in river network studies, with realistic values of the respective parameters. The critical Tokunaga family yields rigorous relations among scaling exponents that have been empirically documented in multiple earlier studies, and serves as a useful analytic and modeling tool for further analysis of river network structure and dynamics. Our results also offer a computationally efficient algorithm of generating self-similar trees with arbitrary parameters (Horton exponents, fractal dimensions, etc.), which facilitates ensemble simulations.  

We represent a stream network that drains a single basin (watershed, catchment) as a rooted binary tree. The basin {\it outlet} (point furthest downstream) corresponds to the tree root, {\it sources} (points furthest upstream) to leaves, {\it junctions} (points where two streams meet) to internal vertices, and {\it links} (stream segments between two successive nodes) to edges. This graph-theoretic nomenclature provides a link to the probability and combinatorics literature on the topic. We assume that all examined trees belong to the space $\cL$ of finite binary rooted planted trees with positive edge lengths \cite{KZ20survey}. Recall that a tree is called {\it planted} if the tree root has degree $1$ (the most downstream link goes to an ocean or another large water body instead of merging with another stream). The space $\cL$ includes the empty tree $\phi$ comprised of the root vertex. We also consider the space $\cT$ of combinatorial projections of trees from $\cL$, that is trees with the same combinatorial structure but no edge lengths.

%=======================================================
\section{Results}

\subsection{Review of Horton laws and implied scaling relationships}%%%%%%%%%%%
The Horton-Strahler orders for river streams have been discussed extensively in the literature and nicely reviewed and summarized in \cite{RIR01}. 
In this section we introduce the orders through the viewpoint of Horton pruning -- this streamlines our exposition and prepares the reader for material that follows. 
For completeness, we also briefly recall Horton laws and their implications.

Consider the map $\cR: \cL \to \cL$ that removes the source links from a tree $\cT$. The Horton-Strahler order of a tree $\cT$ \cite{Horton45,Strahler,Melton59} is the minimal number of Horton prunings that completely eliminates it:
\be\label{eqn:ordTdef}
{\sf ord}(T)=\min\left\{k\ge 0 \,:\, \cR^k(T)=\phi\right\}.
\ee
The Horton pruning and Horton-Strahler orders are illustrated in Fig. \ref{fig:Beaver} (see Appendix~\ref{appdx:HSorder} for details and an alternative computational definition). 

\begin{figure*}[ht]
\includegraphics[width=0.8\textwidth]{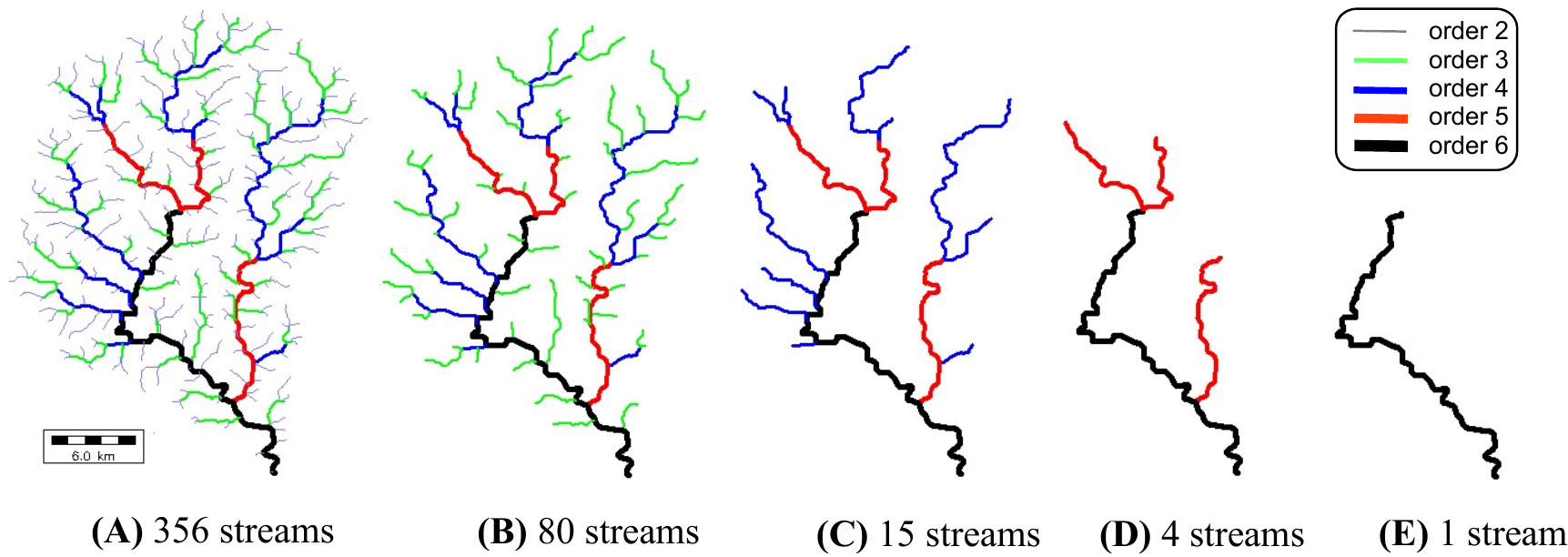}
\caption{\label{fig:Beaver} 
Horton pruning of the stream network of Beaver creek, Floyd County, KY. 
Streams of orders $K=2,\hdots,6$ are shown by different colors (see legend on the right). 
Streams of order $1$ (source streams) are not shown for visual convenience. 
(A) The first Horton pruning, after eliminating streams of order K=1. 
(B)-(E) Second to fifth consecutive Horton prunings. The sixth pruning completely eliminates the network. 
The channel extraction is done using {\sf RiverTools} software (\url{http://rivix.com}).}
\end{figure*}

\begin{figure*}[ht]
\includegraphics[width=0.8\textwidth]{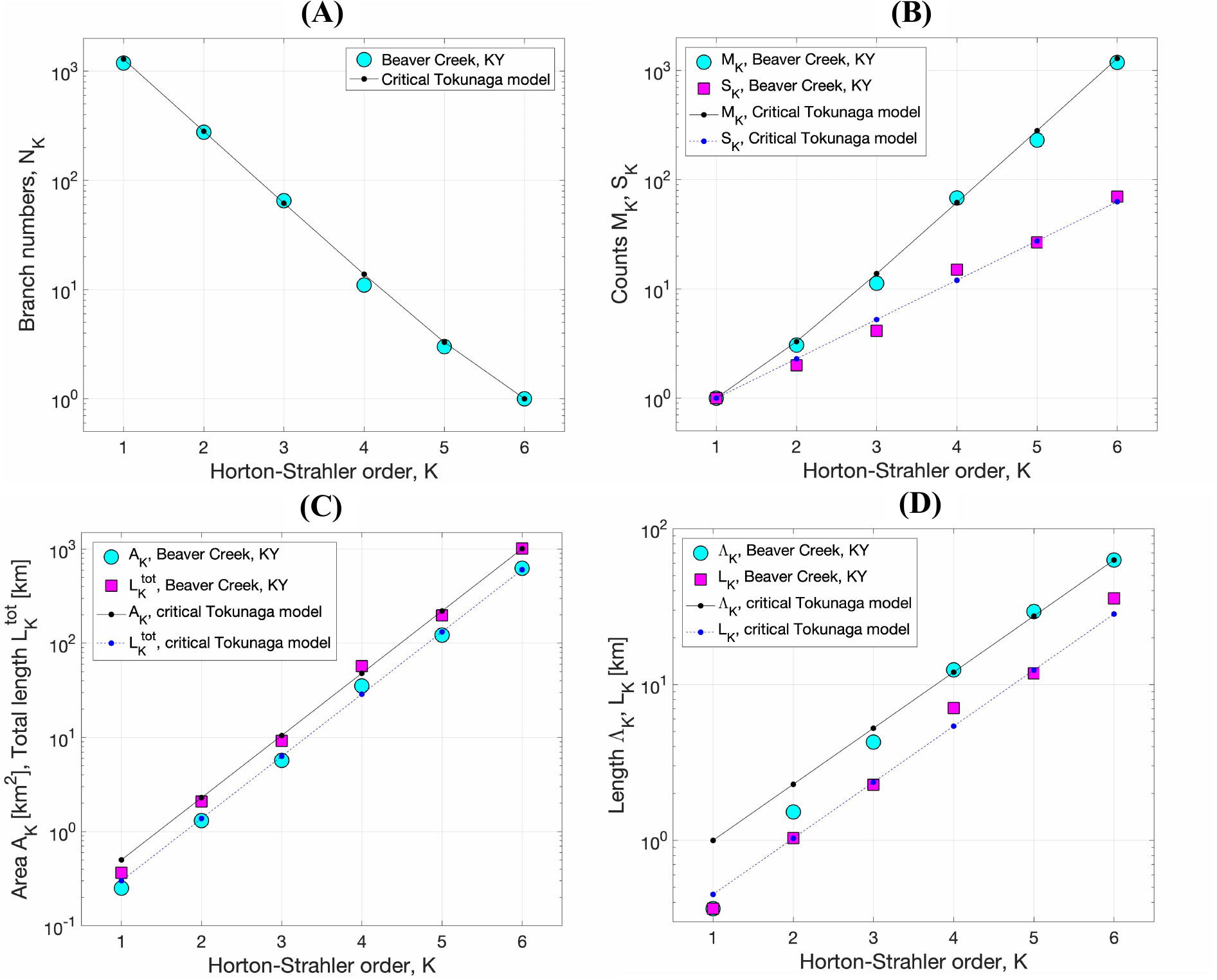}
\caption{\label{fig:Beaver_Horton} 
Critical Tokunaga fit to the Horton laws in the stream network of Beaver creek. The stream network is shown in Fig.~\ref{fig:Beaver}. Symbols correspond to the values of the observed attributes. Lines and dots show the fit by the critical Tokunaga model with  $c=2.3$. 
(A) Stream numbers $N_K$. The model fit is given by \eqref{eqn:Nk}; it has asymptotic slope $-\log_{10}(2c)\approx -0.66$.
(B) Average stream magnitude $M_K$ (cyan circles) and average number of links $S_K$ (magenta squares).
The fit for $M_K$ is given by \eqref{eqn:MNk}; it has asymptotic slope $\log_{10}(2c)\approx 0.66$.
The fit for $S_K$ is given by $c^{K-1}$; it has theoretical slope $\log_{10} c\approx 0.36$.
(C) Average total contributing area $A_K$ (cyan circles) and average total upstream channel length $L_K^{\rm tot}$ (magenta squares).
The fitting lines, according to a combination of equations \eqref{eqn:MNk} and  \eqref{eqn:AvsM}, have theoretical slope $\log_{10}(2c)\approx 0.66$.
(D) Average stream length $L_K$ (magenta squares) and average length $\Lambda_K$ of the longest stream to the divide (cyan circles).
The fitting lines, according to equations \eqref{eqn:HBPL} and \eqref{eqn:Hheight}, have theoretical slope $\log_{10}(c)\approx 0.36$.}
\end{figure*}

Horton's {\it Law of Stream Numbers} \cite{Horton45} postulates a geometric decay of the stream counts $N_K$ of increasing order $K$ with exponent $R_B\geq 2$:
\be\label{eqn:Horton_emp}
N_K~\propto R_B^{-K} \quad\text{ or } \quad \frac{N_K}{N_{K+1}}=R_B.
\ee
The notation $x\propto y$ stands for $x = {\rm Const.}\times y$. The lower bound on $R_B$ follows from the definition of Horton-Strahler orders, since it takes at least two streams of order $K$ to create a single stream of order $K+1$. The value of $R_B$ in large river basins is usually close to $4.5$ \cite{RIR01, Shreve66, GW89, Turcotte_book, PG99, Strahler, Leopold, Tarboton96, GW98, ZZF13, Mesa18}. Figure \ref{fig:Beaver_Horton}A (cyan circles) shows the Horton law for stream numbers in the stream network of Beaver Creek of Fig. \ref{fig:Beaver}; here $R_B \approx 4.6$. 

Horton's {\it Law of Stream Lengths} \cite{Horton45} postulates a geometric growth of the average length $L_K$ of streams of increasing order $K$ with exponent $R_L$:
\be\label{eqn:Horton_LK}
L_K~\propto R_L^K \quad\text{ or } \quad {L_{K+1} \over L_K}=R_L.
\ee
The value of $R_L$ in river networks is around $2.5$ \cite{RIR01}. Figure \ref{fig:Beaver_Horton}D (magenta squares) shows the Horton law for stream 
lengths in the Beaver Creek network of Fig. \ref{fig:Beaver}; here $R_L \approx 2.3$.

Similarly to the laws \eqref{eqn:Horton_emp},\eqref{eqn:Horton_LK} discussed above, 
a geometric scaling of any average stream attribute $Z_K$ with order $K$, 
is also called {\it Horton law}, and the respective exponent is called {\it Horton exponent} \cite{Pec95,NTG97}. 
Horton laws are documented for multiple physical and combinatorial attributes, including upstream area, magnitude (number of upstream sources), total upstream channel length, length of the longest channel to the divide, etc. \cite{RIR01}. 
Figure \ref{fig:Beaver_Horton} shows Horton laws for seven stream attributes of the Beaver Creek that is shown in Fig. \ref{fig:Beaver}. 
We use a convention that the Horton exponent is greater than unity, which always can be achieved by selecting the sign of the exponent K in the Horton law (e.g., Eqs. \eqref{eqn:Horton_emp},\eqref{eqn:Horton_LK}). 

Horton laws imply power-law frequencies of link attributes and power-law relations between the average values of distinct attributes. 
Suppose that stream attributes $Z$ and $Y$ satisfy Horton's law with Horton exponents $R_Z$ and $R_Y$, respectively. 
Using each of the laws to express the channel order $K$ and equating these expressions, we find
\be\label{eqn:HL1}
Z_K~\propto~Y_K^\alpha,\quad\text{ with } \quad \alpha = \frac{\log R_Z}{\log R_Y}.
\ee
Equation \eqref{eqn:HL1} is a punctured (by discrete orders) version of a power-law relation $Z\propto Y^\alpha$  that is abound among hydrologic quantities. 
A well-studied example is Hack's law that relates the length $L$ of the longest stream in a basin to the basin area $A$ via $L\propto A^{\bf h}$ with ${\bf h} \approx 0.6$ \cite{Hack57, Rigon+96}. Equation \eqref{eqn:HL1} suggests that the parameter ${\bf h}$ is expressed via the exponents $R_L$ and $R_A$ of the Horton laws for length $L$ and area $A$ as:
\be\label{eqn:hRlRa}
{\bf h}=\frac{\log R_L}{\log R_A}.
\ee

Next, consider the value $Z_{(i)}$ of an attribute $Z$ calculated at link $i$ in a large basin. Assuming Horton laws for $N_K$ and $S_K$ with exponents $R_B$ and $R_S$, respectively, and considering a limit of an infinitely large basin we approximate the distribution of link attributes as (see Appendix~\ref{appdx:linkatrib})
\be\label{eqn:HL2}
\#\{i:Z_{(i)}\ge z\}~\propto~ z^{-\beta},\quad\beta = \frac{\log R_B - \log R_S}{\log R_Z}.
\ee

Such power laws are documented for the upstream contributing area, length of the longest channel to the divide, water discharge, or energy expenditure. For example, analyses of river basins (e.g., \cite{MRR+96,TBR88,RI92}) extracted from digital elevation models (DEM's) suggest
\be\label{intro:powerA}
\#\{i:A_{(i)}\ge a\} \propto a^{-\beta_A},\quad \beta_A \approx 0.43
\ee
and
\be\label{intro:powerL}
\#\{i:\Lambda_{(i)}\ge l\} \propto l^{-\beta_{\Lambda}},\quad \beta_\Lambda \approx 0.8,
\ee
where $A_{(i)}$ is the area upstream of link $i$ and $\Lambda_{(i)}$ is the distance from link $i$ to the furthest source (or, equivalently, to the basin divide) measured along the channel network. 

Horton laws (e.g. Eqs. \eqref{eqn:Horton_emp},\eqref{eqn:Horton_LK}) and the implied scaling relations (Eqs. \eqref{eqn:HL1},\eqref{eqn:HL2}) provide key observational constraints in modeling river stream networks \cite{RIR01,RRR+93,Mesa18,TBR88}. 
Our work explains the appearance of Horton laws in terms of tree self-similarity and offers a parametric toolbox for the analysis and modeling of river networks
and other branching structures that exhibit such scaling relations.

\subsection{Tree self-similarity and Tokunaga sequence}%%%%%%%%%
We introduce the concept of self-similarity for random trees that encompasses the existing definitions and satisfies practical intuitive expectations. The proposed definition applies to a distribution of trees from a suitable space such as $\cT$ or $\cL$ and combines two fundamental properties -- {\it coordination} and {\it Horton prune-invariance}. 

Coordination means that the (random) structure of a river basin is determined by its order. 
For example, a basin with outlet of order three and a sub-basin of order three within a basin of order nine have, statistically, the same structure. 
This assumption is at the heart of analyses based on the Horton-Strahler orders; it has been imposed, explicitly or implicitly, in the mainstream studies of river networks \cite{Horton45,RIR01,Shreve66,DBE94,Pec95,PG99,Tarboton96}. 
A distribution that satisfies the coordination property is called coordinated.
We refer to \cite{KZ20survey} for a measure-theoretic definition of coordination. 

Horton prune-invariance formalizes the expectation that the scaling laws of hydrology are (by and large) independent of the data resolution. The Horton pruning $\cR$ is a natural model for the change of resolution in a river network. Indeed, better observations lead to detecting smaller streams, which increases the basin order. Pruning a basin by order is roughly equivalent to decreasing the resolution of stream detection. The Horton prune-invariance requires that the statistical structure of trees remains the same after zooming in or out. 

\begin{Def}[Self-similar tree]\label{def:SST} 
A coordinated distribution $\mu$ on the space $\mathcal{T}$ of combinatorial 
trees is called {\it self-similar} if it is invariant with respect to 
Horton pruning \cite{BWW00,KZ16}:
\begin{equation}\label{eqn:ssm}
\mu\left(\mathcal{R}^{-1}(T)|T\ne\phi\right)=\mu(T)
\quad\text{for any}\quad T\in\mathcal{T}.
\end{equation}
\end{Def}
Recall that $\phi$ denotes an empty tree. Equation \eqref{eqn:ssm} states that, for any non-empty tree $T$, the total probability assigned by $\mu$ to all trees that result in $T$ after pruning -- these trees are collectively denoted by $R^{-1}(T)$ -- is the same as the probability of $T$. Informally, consider a forest of trees generated by measure $\mu$, where each tree $T$ occurs multiple times according to its probability $\mu(T)$. The forest is self-similar if after pruning each tree by $\cR$ we obtain the same forest. This definition can be extended to trees with edge lengths from space $\cL$; see Def. 9 in \cite{KZ20survey}. 
In that case, we allow the edge lengths to scale after pruning by a multiplicative scaling constant $\zeta > 0$. 
We use a conventional abuse of terminology by saying that a tree $T$ is self-similar meaning that $T$ is a random tree drawn from a self-similar distribution $\mu$.

The measure-theoretic Definition \ref{def:SST} might be not appealing for practical analyses that oftentimes involve only a handful of finite basins. A bridge from this definition to easily computed stream statistics is provided by Theorem \ref{thm:TokSeq} (see Appendix~\ref{appdx:sstTokHL}). It shows that the expected value $T_{i,j}$ of the number $n_{i,j}$ of streams of order $i$ that merge with a randomly selected stream of order $j$ in a self-similar tree is a function of the difference $j-i$ (and not individual values of $i$ and $j$). Accordingly, each self-similar measure $\mu$ corresponds to a unique non-negative sequence of {\it Tokunaga coefficients} 
\begin{equation}\label{tokuind2}
T_{i,i+k}=T_k={\sf E}_\mu [n_{i,i+k}] \quad \text{for all}\qquad i,k>0,
\end{equation}
where ${\sf E}_\mu$ denotes the mathematical expectation with respect to $\mu$. 
Moreover, we show (see Cor.~\ref{cor:PruneVsToeplitz} in Appendix~\ref{appdx:sstTokHL}) that the Toeplitz property $T_{i,i+k}=T_k$ and Horton prune-invariance of Eq. \eqref{eqn:ssm} are equivalent for coordinated measures (i.e., both hold or do not hold at the same time). 
Hence, the Tokunaga coefficients $T_k$ provide a fundamental parameterization of a self-similar tree and constitute the main tool of respective applied analyses. 

Our Definition \ref{def:SST} of tree self-similarity unifies the alternative definitions used in the literature. Burd et al. \cite{BWW00} define self-similarity in Galton-Watson trees as the Horton prune-invariance; this is a special case of our definition since the Galton-Watson trees are coordinated \cite{KZ20survey}. Peckham \cite{Pec95} and Newman et al. \cite{NTG97} define self-similarity as the Toeplitz property for Tokunaga coefficients; this is equivalent to our definition in coordinated trees (Corollary \ref{cor:PruneVsToeplitz}). 
The coordination assumption is further justified in \cite{KZ20survey} by showing that 
the Toeplitz property alone, without coordination, allows for a multitude of obscure measures that are hardly useful in practice. 
Gupta and Waymire \cite{GW89} and Peckham and Gupta \cite{PG99} suggested a concept of {\it statistical self-similarity} that requires a random stream attribute $Z$ to have distribution that scales with order. It can be shown (Sect. 7 in \cite{KZ20survey}) that (i) statistical self-similarity for some attributes (e.g., for any discrete attribute) may only hold asymptotically, and (ii) multiple attributes, including stream length, magnitude, and total basin length, are statistically self-similar in a limit of infinitely large basin that is self-similar according to our Definition \ref{def:SST}.

\subsection{Horton laws for stream numbers, magnitudes in self-similar trees}%%%%%%%%%

We now capitalize on the concept of tree self-similarity introduced above to
establish a key emergent property of self-similar trees -- Horton laws for stream numbers and magnitudes, conveniently parameterized by the Tokunaga sequence.

Consider the mean number $$\cN_i[K] = {\sf E}_\mu[N_i(T)\,|\, {\sf ord}(T) = K]$$ of streams of order $i$ in a basin of order $K$ and the mean magnitude (number of upstream sources) $M_i$ of a stream of order $i$. Consider also the generating function $T(z)=\sum\limits_{k=1}^\infty T_k z^k$ of the Tokunaga coefficients and define
\be\label{eqn:tofz}
\hat{t}(z) = -1 + 2z + T(z).
\ee
\begin{thm}[Horton law for stream numbers, magnitudes in self-similar tree \cite{KZ20survey,KZ16}]\label{thm:M}
Consider a self-similar tree $T$ with Tokunaga sequence $T_k$ and suppose that 
\be\label{eqn:limsupTk}
\limsup\limits_{k \to \infty} T_k^{1/k}<\infty.
\ee
Then, the stream numbers $\cN_i[K]$ and the magnitudes $M_i$ obey Horton laws:
\be\label{eqn:Nk}
\lim_{K\to\infty}\left(\cN_1[K]\,R_B^{-K}\right) = M ~<\infty,
\ee
\be\label{eqn:MNk}
\lim_{i\to\infty}\left(M_i\,R_M^{-i}\right) = M ~<\infty.
\ee
The Horton exponents are given by 
\be\label{eqn:RbRmW0}
R_B=R_M=1/w_0,
\ee
where $w_0$  is the only real root of the function $\hat{t}(z)$ in the interval $(0,1/2]$
and $M$ is a positive real constant given by
\be
M =- \frac{1}{w_0}\lim_{z\to w_0}\frac{z(z-w_0)}{\hat{t}(z)}.
\ee
\end{thm}
The proof is given in Appendix~\ref{appdx:proof1}.

Theorem \ref{thm:M} shows that the Horton laws for mean stream numbers $\cN_i[K]$ and magnitudes $M_i$ hold in almost any self-similar tree, more specifically -- in a tree with an arbitrary Tokunaga sequence $T_k$. The only restriction of Eq. \eqref{eqn:limsupTk} prohibits super-exponential growth of $T_k$, such as $T_k = k!$ or $T_k = k^k$. 
The theorem establishes a strong form of the Horton law (Eqs. \eqref{eqn:Nk},\eqref{eqn:MNk}), which implies a weaker version that is often reported in applied literature:
\begin{eqnarray}\label{eqn:meanHorton}
\lim\limits_{K\to\infty}\frac{\cN_i[K]}{\cN_{i+1}[K]}&=&R_B ~~~\text{for any }i,\text{ and }\nonumber\\
\lim\limits_{i\to\infty}\frac{M_{i+1}}{M_i}&=&R_M.
\end{eqnarray}

Theorem \ref{thm:M} emphasizes the existence of a multitude of self-similar measures with the same Horton exponent. Assume we fix $R_B$ and hence the root $w_0$ of $\hat{t}(z)$ according to Eq. \eqref{eqn:RbRmW0}. Equation \eqref{eqn:tofz} readily asserts that there is an infinite number of Tokunaga sequences that correspond to an arbitrary $w_0$ within $(0,1/2]$. For example, if $R_B = 4$, then $w_0 = 1/4$ and one needs $T(z) = 1/2$. This can be achieved by selecting any of $T_k = \{2,0,\hdots\}$, $\,\{1,4,0,\hdots\}$, $\,\{0,8,0,\hdots\}$, $\,\{2^{k-1}\}$, etc., where ``$\hdots$'' denotes trailing zeros.

We observe that the Horton law of stream numbers in Theorem \ref{thm:M} (Eqs. \eqref{eqn:Nk}) is an asymptotic statement, different from the ideal Horton law for stream numbers \eqref{eqn:Horton_emp} which is commonly used in the literature. This is not a mathematical peculiarity -- the {\it ideal} Horton law is merely an approximation to the actual behavior of stream counts. Its approximate nature is not related to the finite size of the observed basins -- the ideal Horton law rarely holds in theoretical trees of arbitrarily large size.  Formally, we show in Appendix~\ref{appdx:eHorton} that the ideal Horton law for stream numbers in a self-similar tree holds if and only if $T_k = 0$ for $k > 1$. Realistically, Horton laws are asymptotic statements of different strength. For example, the strongest form of Horton law for stream numbers is that of Eq. \eqref{eqn:Nk}, which implies a weaker version of Eq. \eqref{eqn:meanHorton}. Accordingly, the power relations among different stream attributes \eqref{eqn:HL1} and power-law frequencies of link attributes \eqref{eqn:HL2} that we have derived from the ideal Horton law of Eq. \eqref{eqn:Horton_emp} remain heuristic. A formal analysis based on actual Horton laws (like those in Eqs. \eqref{eqn:Nk} and \eqref{eqn:MNk}), which will be presented elsewhere, confirms the results of Eqs. \eqref{eqn:HL1} and \eqref{eqn:HL2} and reveals additional solutions with oscillatory tail behavior. 

The Horton laws for other stream attributes may or may not hold depending on additional assumptions about $T_k$ and other details of basin organization. A comprehensive treatment is possible using the generating function approach outlined in 
Appendix~\ref{appdx:asymp}. Most importantly, further analysis often requires specifying a concrete self-similar distribution, not only its Tokunaga sequence $T_k$. Below we examine a particularly useful family of distributions.

\subsection{Random Attachment Model (RAM) of self-similar trees}%%%%%%%%%%%
According to Theorem \ref{thm:TokSeq} (Appendix~\ref{appdx:sstTokHL}), every self-similar measure corresponds to a unique Tokunaga sequence $T_k$. At the same time, a multitude of self-similar measures can be constructed for a given Tokunaga sequence. Here we introduce a particularly symmetric random tree (tree distribution) for a given Tokunaga sequence and establish its key properties. We use Poisson attachment construction within exponential segments; this ensures that the link lengths have exponential distribution and the attachment of streams of lower orders to a given stream of a larger order is done in uniform random fashion. We refer to this construction as {\it Random Attachment Model} (RAM).

The RAM specifies a tree distribution on $\cL$ by a non-negative Tokunaga sequence $T_k$, the order distribution $\pi_K = {\sf P}({\sf ord}(T) = K)$, and the distribution of stream lengths. The model assumes that the lengths of streams of order $j$ are independent exponential random variables with rate $\lambda_j$. Hence, the model is specified by three non-negative sequences:
$$\{T_k\}, ~~\{\lambda_j\},~\text{ and }~\{\pi_K\}.$$
Importantly, the probabilities $\pi_K$ and rates $\lambda_j$ do not affect the combinatorial structure of a tree of a given order, which is completely specified by the Tokunaga coefficients $T_k$. 

A random tree is constructed in a hierarchical fashion, starting from the stream of the highest order and adding side-tributaries of consecutively smaller orders. The tree order $K$ is selected according to the distribution $\pi_K$. 
At the first step we generate the main stream that will have order $K$ in the final tree; its length is an exponential random variable with rate  $\lambda_K$. At each of the remaining $K-1$ steps, we add streams of lower orders to the existing tree by a Poisson attachment procedure. The streams added at step $m$ will have order $i(m) = K-m+1$ in the final tree. The lengths of the newly added streams are independent exponential random variables with rate $\lambda_{i(m)}$. The new streams are added in two steps. First, we consider the existing tree as a one-dimensional metric space (union of link segments) and generate a collection of points on this space according to a homogeneous Poisson process. The process intensity depends on the order of a link within the final tree. Specifically, within every link added at step $K-j+1$ (that will have order $j$ in the final tree) the Poisson intensity is $\lambda_j T_{j-i(m)}$. A single stream is then attached to each Poisson point. Second, we add two new streams to each source stream of the current tree (except the sources just added during this step). The first part of this procedure (Poisson attachment) ensures that the tree has Tokunaga coefficients $T_k$, and the second part (adding stream pairs) increases the tree order by one at each step.

\begin{figure*}[ht] %[p] [t]
\includegraphics[width=0.8\textwidth]{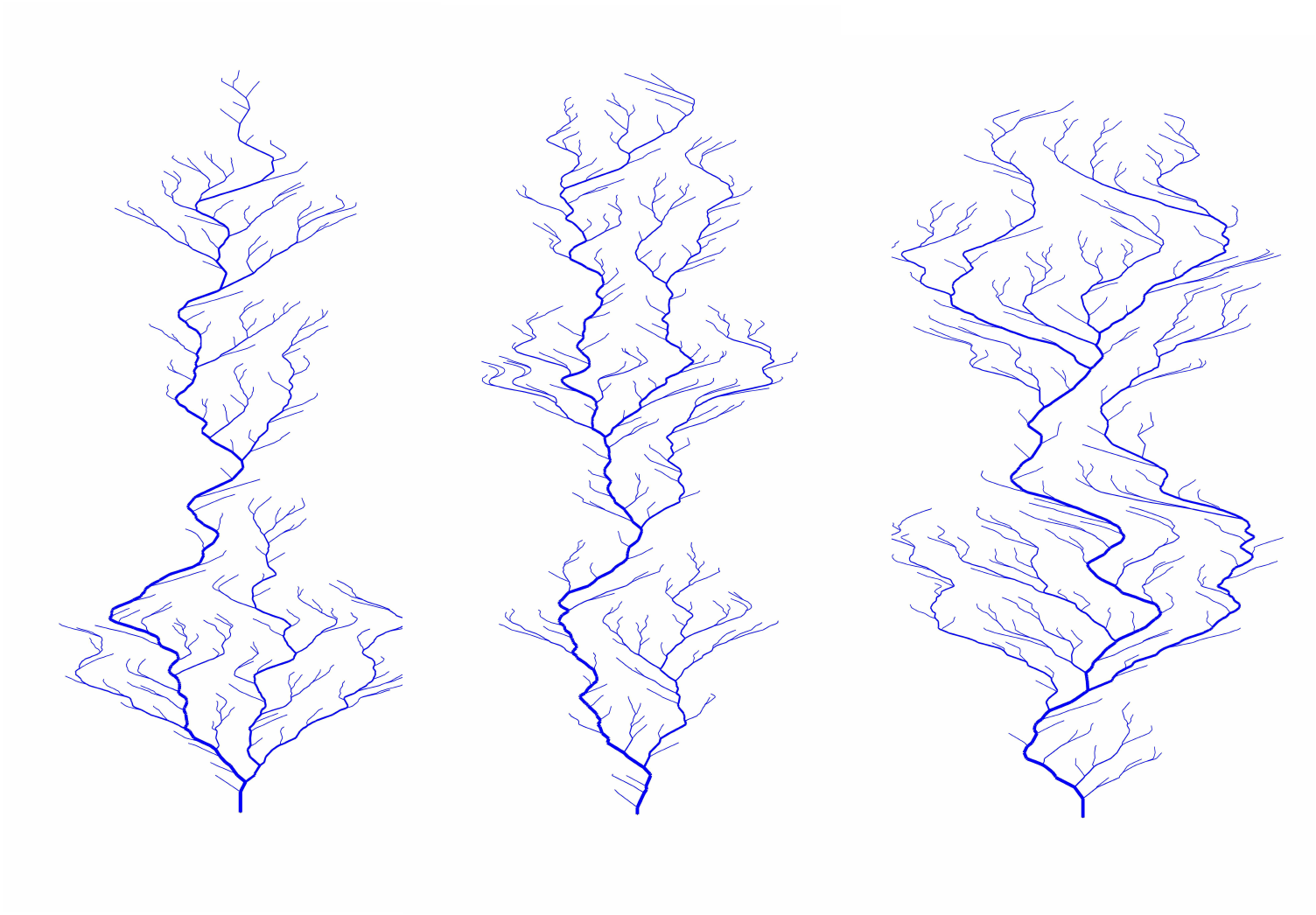}
\caption{Examples of critical Tokunaga trees.
The trees are generated by the critical Tokunaga process with $c=2.3$
and Horton-Strahler order $K=5$. 
The line width is proportional to the contributing area. The figure accurately represents the tree combinatorial structure; the edge lengths are scaled for a better planar embedding.}
\label{fig:HBP_ex}
\end{figure*}

The trees generated by RAM can be equivalently represented as trajectories of a continuous-time multitype {\it Hierarchical Branching Process}, with time evolving from the root upstream and member types corresponding to the stream orders. This approach, explored by the authors in \cite{KZ20survey}, yields the joint distribution of the orders $K_1 < K_2$ of subtrees that share a common root of order $K$:
\be
\label{eqn:joint}
{\sf P}\left(K_1=j,K_2=m|K\right)
=
\begin{cases}
S^{-1}_{K}&\!\!\text{if }j=m=K-1,\\
T_{K-j}S^{-1}_{K}&\!\!\text{if }j<m=K.
\end{cases}
\ee
We now use this result to propose a computationally efficient recursive construction of RAM trees. A tree of order $1$ is a stream of exponential length with rate $\gamma$. To create a tree of order $K > 1$ we first generate a link of exponential length with rate $\lambda_K S_K$, where
\begin{equation}\label{eqn:Ck}
S_K = 1+\sum_{i=1}^{K-1}T_i.
\end{equation}
To this link we attach two conditionally (conditioned on the order $K$) independent trees whose orders are drawn from Eq.\eqref{eqn:joint}. Each of these trees is generated according to the same recursive procedure. This algorithm generates trees with up to $10^6$ edges within seconds, providing a flexible computational framework for ensemble simulations based on independent statistical realizations of a tree with fixed parameters. Examples of RAM stream networks are shown in Fig. \ref{fig:HBP_ex}.

Another useful result of the branching process theory establishes the necessary and sufficient conditions for a RAM tree to be self-similar according to Definition \ref{def:SST}. These conditions explicitly parameterize the probabilities $\pi_K$ and stream length rates $\lambda_j$ for an {\it arbitrary} Tokunaga sequence $T_k$. This emphasizes the richness of self-similar family. 

\begin{thm}[Self-similar RAM; Thm. 11 in \cite{KZ20survey}]\label{thm:sstRAM} 
Suppose $T$ is a random tree generated by the RAM with parameters $\{T_k\}$, $\{\lambda_j\}$, and $\{\pi_K\}$. 
Then $T$ is a coordinated tree. Tree $T$ is self-similar with scaling constant $\zeta>0$ (see Definition  \ref{def:SST} and its discussion) if and only if 
\begin{eqnarray}\label{eqn:sstRAM}
\pi_K&=& p(1-p)^{K-1} ~~(K \geq 1) ~~~\text{ and }\nonumber\\
\lambda_j&=&\gamma\,\zeta^{1-j} ~~~(j \geq 1)
\end{eqnarray}
for some parameters $p \in (0,1)$, $\gamma>0$ and $\zeta>0$ (and any Tokunaga sequence $T_k$).
\end{thm}

\begin{cor}[Horton law for the stream lengths]\label{cor:HBPL} 
Consider a self-similar RAM tree with parameters given by Eq. \eqref{eqn:sstRAM}. Then the average length $L_j$ of a stream of order $j$ satisfies
\be\label{eqn:HBPL}
L_j R_L^{-j} =\frac{1}{\zeta\gamma}<\infty \quad \text{ with }\quad R_L=\zeta.
\ee
\end{cor}
The proof is given in Appendix~\ref{appdx:ctcProof}.
We notice that the Horton law holds here in an exact form, without a limit in order $j$.

We show below that the two well-known properties of self-similar river basins -- fractal dimension and Horton law for the longest stream length -- formally hold in a self-similar RAM tree. 

\begin{thm}[Fractal dimension of a self-similar tree]\label{thm:FractalD} 
Consider a self-similar RAM tree with Tokunaga sequence $T_k$ and other parameters given by Eq. \eqref{eqn:sstRAM}. Let $w_0$ be the only real root of the function $\hat{t}(z)$ (Eq. \eqref{eqn:tofz}) in the interval $(0, 1/2]$.  Then, the fractal dimension of the tree in the limit of infinite order and after a suitable length rescaling, is given by
\be\label{eqn:FDlogs}
{\bf d}=\max\{1,{\bf d}_0\},\quad{\bf d}_0 = -\frac{\log w_0}{\log\zeta} = \frac{\log R_B}{\log R_L}.
\ee
\end{thm}
The proof is given in Appendix~\ref{appdx:fracD}.\\

\begin{figure*}[ht]
\includegraphics[width=0.8\textwidth]{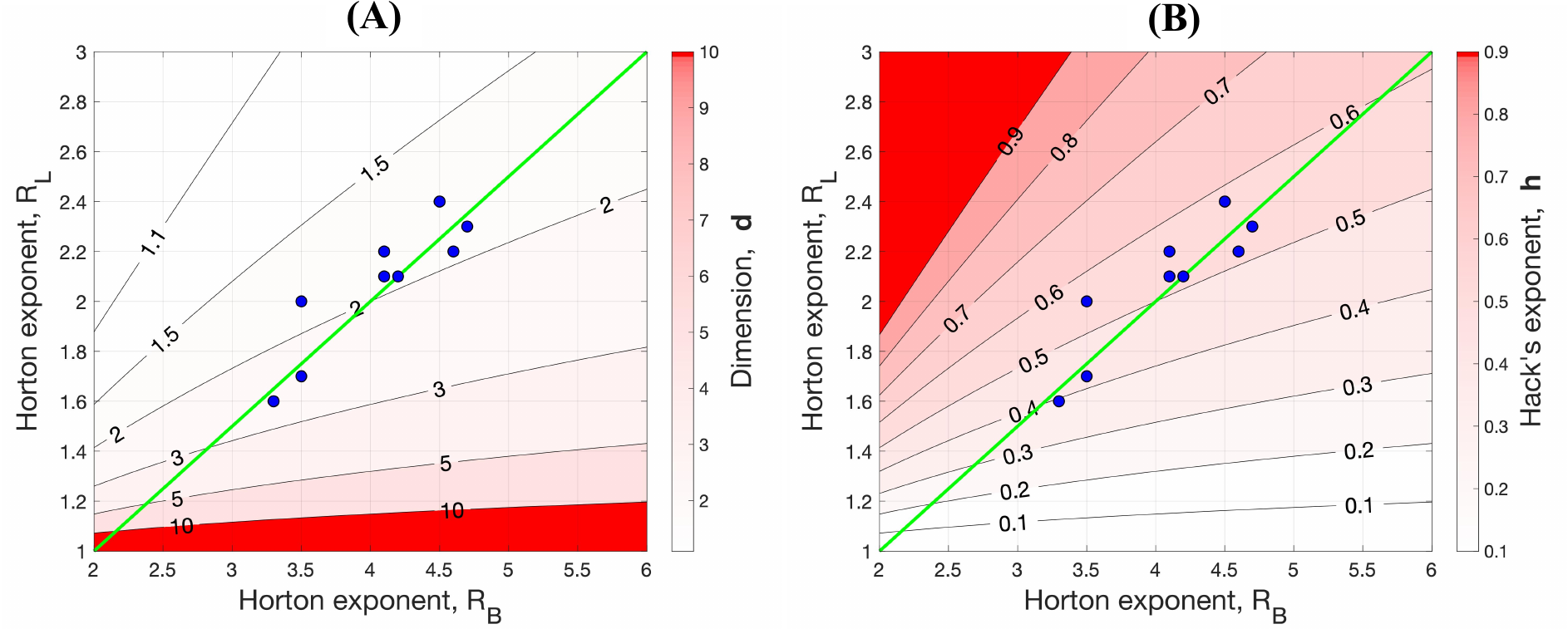}
\caption{\label{fig:d} 
Fractal dimension ${\bf d}$ (panel {\bf A}) of equation \eqref{eqn:FDlogs} and
Hack's exponent ${\bf h} = {\bf d}^{-1}$ (panel {\bf B}) of a self-similar RAM tree in the limit of infinite size as a function of the Horton exponents $R_B$ and $R_L$.
Selected levels of ${\bf d}$ and ${\bf h}$ are shown by marked black lines. Green thick lines correspond to the critical Tokunaga process (Definition \ref{def:TokTree}, Eq. \eqref{eqn:Tok}) for which $R_B = 2R_L$. 
Blue dots depict the pairs $(R_B,R_L)$ estimated in nine real river basins by \cite{TBR88}, see also \cite[Table~2.1]{RIR01}.
}
\end{figure*}

Equation \eqref{eqn:FDlogs} coincides with the expression first obtained by La Barbera and Rosso \cite{LR89} using a heuristic assumption of a basin with an ideal Horton law of stream numbers. Figure \ref{fig:d}A shows a map of ${\bf d}$ as a function of the Horton exponents $R_B$ and $R_L$. 

\begin{thm}[Horton law for the length of the longest stream]\label{thm:HortonLawLambda} 
Consider a self-similar RAM basin with parameters given by Eq. \eqref{eqn:sstRAM}. Let $\Lambda_k$ denote the average length of the longest stream in a basin of order $k$. Then
\be\label{eqn:Hheight}
\lim\limits_{k \to \infty} \Lambda_k R_\Lambda^{-k}={\rm Const.}<\infty \quad \text{ with }\quad  R_\Lambda=\zeta.
\ee
\end{thm}
The proof is given in Appendix~\ref{sec:HL}.

\subsection{Critical Tokunaga tree and emergent scaling relations}%%%%%%%%%%%%
Observations on river networks have supported a basic constraint that the link length distribution is independent of the position of the link within a basin \cite{RIR01,TBR89}. This motivates one to describe a family of trees that respect this property. Surprisingly, this leads to a one-parameter family of {\it critical Tokunaga trees} that satisfy multiple additional symmetries and include the celebrated Shreve model as a special case. 

The length of a link of order $K$ in the RAM model is an exponential random variable with rate $\lambda_K S_K$, which is a direct consequence of using Poisson attachment along exponentially distributed streams. The order-independent link length implies $\lambda_K S_K = {\rm Const.}$ Using the general form of $\lambda_K$ in a self-similar RAM tree \eqref{eqn:sstRAM}, one can select $\gamma$ such that $S_K = \zeta^{K-1}$. This corresponds to the unique form of the Tokunaga sequence in a self-similar RAM tree with identically distributed link lengths:
\be\label{eqn:Tok}
T_k=(c-1)\,c^{k-1} ~~~(k \geq 1),\qquad c=\zeta \geq 1.
\ee

\begin{Def}[Critical Tokunaga tree]\label{def:TokTree}
A self-similar RAM with $p={1 \over 2}$ and Tokunaga sequence of Eq. \eqref{eqn:Tok} is called critical Tokunaga tree or critical Tokunaga model. 
By Theorem \ref{thm:sstRAM}, for this model we have $\pi_K=2^{-K}$ and $\lambda_j= \gamma \, \zeta^{1-j}$.
\end{Def}

Figure \ref{fig:HBP_ex} shows three examples of critical Tokunaga trees of order $K = 5$ with parameter $c = 2.3$, which gives a close approximation to the observed river networks (see Table \ref{tab:cTok}). Equation \eqref{eqn:Tok} is a special case of the two-parameter sequence $T_k = ac^{k-1}$ introduced by Eliji Tokunaga \cite{Tok78} to approximate river basin branching; this sequence has been examined in detail in \cite{Pec95,Turcotte_book,NTG97,ZZF13,KZ16}. The one-parameter sequence of Eq. \eqref{eqn:Tok} appears in the random self-similar network (RSN) model of Veitzer and Gupta \cite{VG00}, which uses a purely combinatorial algorithm of recursive local replacement of the network generators to construct random trees. A theoretical underpinning of this constraint is revealed via the prism of branching process analysis. Kovchegov and Zaliapin \cite{KZ18} have shown that a random tree $T$ generated by the critical Tokunaga model is {\it critical} and {\it time-invariant} in both combinatorial and metric forms \cite{KZ20survey}. In particular, the condition $p = 1/2$ is necessary and sufficient for criticality. Moreover, the Geometric Branching Process (that generates the combinatorial part of a RAM tree) is time invariant if and only if it corresponds to the critical Tokunaga model. Recall that {\it criticality} means that a branching process has unit average population size after an arbitrary but fixed time advancement (in both discrete and continuous versions). {\it Time-invariance} means that the frequency of orders of subtrees that survive after a given time advancement coincides with the initial order distribution. 

It is natural to assume that the {\it local contributing area} of a link (area that contributes to the link directly, and not via its descendant joint) is a function of the link length. This allows us to examine the average total contributing area $A_i$ of a stream of order $i$. In particular, the order-independent link lengths imply order-independent local areas. The following result establishes Hack's law in a critical Tokunaga tree.

\begin{thm}[Hack's law in a critical Tokunaga tree]\label{thm:HacksLawTok} 
Consider a critical Tokunaga tree (Definition \ref{def:TokTree}). Then the average lengths $\Lambda_i$ of the longest stream and the average total contributing area $A_i$ of a basin are related as
\begin{eqnarray}\label{eqn:HacksLawTok}
\Lambda_i &\sim& {\rm Const.}\times \big(A_i\big)^{\!{\bf h}}, 
\text{ where }\nonumber\\
{\bf h}&=&{\bf d}^{-1} \!\!= -{\log\zeta \over \log w_0} = \frac{\log R_L}{\log R_B}.
\end{eqnarray}
\end{thm}
The proof is given in Appendix~\ref{appdx:ctcProof}.\\

The Hack's law of Eq. \eqref{eqn:HacksLawTok} also holds in more general self-similar RAM trees (which may not be critical Tokunaga) as is shown in Appendix~\ref{appdx:fracD}. Figure \ref{fig:d}B shows a map of ${\bf h}$ as a function of the Horton exponents $R_B$ and $R_L$. The critical Tokunaga case corresponds to $R_B = 2R_L = 2c$ and hence ${\bf d} = \log_c(2c)$ and ${\bf h} = \log_{2c}c$; this case is depicted by green line in Fig. \ref{fig:d}.

Combining our results, we obtain the following summary of the Horton exponents in a critical Tokunaga tree.
\begin{cor}[Horton exponents in a critical Tokunaga tree]\label{cor:cRbRmRaRsRl} 
The Horton exponents in the critical Tokunaga model are given by 
\be\label{eqn:cRbRmRaRsRl}
2c = R_B = R_M = R_A > R_S = R_L = R_\Lambda = c.
\ee
\end{cor}
The proof is given in Appendix~\ref{appdx:ctcProof}.

Corollary \ref{cor:cRbRmRaRsRl} reveals that the essential Horton exponents in critical Tokunaga trees only assume two distinct values ($c$ and $2c$). The inequality $R_S < R_B$, which is a part of Eq. \eqref{eqn:cRbRmRaRsRl}, has been conjectured by Peckham \cite{Pec95} for trees with a well-defined Tokunaga sequence.

Notably, the critical Galton-Watson process with exponential edge lengths \cite{Pitman}, which is equivalent to Shreve's random topology model after conditioning on the basin magnitude, is a special case of critical Tokunaga model with $c = 2$ \cite[Thm.~15]{KZ20survey}. 
In other words, the critical Tokunaga model offers a natural extension of the critical binary Galton-Watson process to a similarly versatile family of processes with a wide range of Horton exponents, fractal dimensions, Hack's exponents, and other parameters. As such, this model may be useful for multiple fields beyond hydrology.

Results of Chunikhina \cite{EVC18,EVCthesis} imply that the critical Tokunaga model with $c = 2$ maximizes the entropy rate among the trees that satisfy the Horton law of stream numbers, and that the critical Tokunaga model with a fixed $c$ maximizes the entropy rate among the trees that satisfy the Horton law for stream numbers with $R_B = 2c$.

Some additional scaling properties of the critical Tokunaga trees are collected in Appendix~\ref{appdx:cTok}.

\subsection{Critical Tokunaga model closely fits observations}%%%%%%%%%%%%
The critical Tokunaga model provides a very close fit to the data and scaling relations reported in river studies over the past decades. Table \ref{tab:cTok} summarizes the values of the key scaling exponents in the critical Tokunaga model and compares them with exponents in river network observations and the well-established OCN model \cite{RRR+93}. The table uses the results of Corollary \ref{cor:cRbRmRaRsRl}, Eq. \eqref{eqn:cRbRmRaRsRl} (Horton exponents), Theorem \ref{thm:FractalD}, Eq. \eqref{eqn:FDlogs} (fractal dimension ${\bf d}$), Theorem \ref{thm:HacksLawTok}, Eq. \eqref{eqn:HacksLawTok} (Hack's exponent ${\bf h}$), and Eq. \eqref{eqn:HL2} (scaling exponents $\beta_\Lambda$ and $\beta_A$).

The critical Tokunaga model fit to the observed data is further illustrated in the Beaver Creek basin of Fig. \ref{fig:Beaver}. Figure \ref{fig:Beaver_Horton} shows seven Horton laws fit by the critical Tokunaga model with $c = 2.3$. Specifically, we consider the following stream attributes averaged over streams of order $K = 1,\hdots,6$: the stream number $N_K$ (panel A), the average magnitude $M_K$ and the average number $S_K$ of links in a stream (panel B), the average total contributing area $A_K$ and the average total upstream channel length $L^{tot}_K$ (panel C), and the average stream length $L_K$ and the average length $\Lambda_K$ of the longest channel to the divide (panel D). The fitting lines correspond to the critical Tokunaga model predictions, which impressively agree with observations of all examined stream attributes (see figure caption).

%=======================================================
\section{Discussion}
A solid body of observational, modeling, and theoretical studies connects Horton laws, power-law distributions of and power-law relations among river stream attributes 
to the self-similar structure of stream networks \cite{RIR01,BWW00,Pec95,GW89,GTD07,Turcotte_book,PG99,Tarboton96,GW98,Mesa18,VG00,DR00,GCO96,Cetal98,PT00}. We suggest a rigorous treatment of the appearance and parameterization of Horton laws in river networks using a recently formulated theory of random self-similar trees \cite{KZ20survey}. The proposed framework unifies the existing results and contributes to explaining the ubiquity of Horton laws in dendritic systems of arbitrary origin. 

The main technical contribution of our work is a rigorous treatment of the appearance of Horton laws in self-similar trees (Eqs. \eqref{eqn:Nk},\eqref{eqn:MNk},\eqref{eqn:meanHorton},\eqref{eqn:HBPL},\eqref{eqn:Hheight}). We show that the two fundamental properties -- {\it coordination} and {\it Horton-prune invariance} -- necessarily lead to the Horton laws for stream numbers and magnitudes (Theorem \ref{thm:M}). Additional mild assumptions, like those in the Random Attachment Model (RAM), yield the Horton laws for multiple other attributes (Theorem \ref{thm:HortonLawLambda}; Corollaries \ref{cor:HBPL},\ref{cor:cRbRmRaRsRl}), which in turn imply basin fractal dimension (Theorem \ref{thm:FractalD}), Hack's law (Theorem \ref{thm:HacksLawTok}), and other power-law scaling relations (Eqs. \eqref{eqn:HL1},\eqref{eqn:HL2}). Our results can be easily extended to other stream attributes such as stream slope, width, depth and velocity, which are known to be proportional to a power of the upstream magnitude \cite{RIR01,GW89}. The developed framework may also facilitate analysis of the width function \cite{LF07} or scaling of hydrologic fluxes \cite{GTD07,GW90} in self-similar basins. Such analyses can be done either analytically, or using ensemble simulation that is facilitated by a fast simulation algorithm for RAM trees.

The self-similarity is defined here (Definition \ref{def:SST}) as invariance of a coordinated tree distribution with respect to the operation of Horton pruning, which is in accord with the empirical and modeling evidence of the past decades \cite{BWW00,Pec95,NTG97,Melton59,KZ16,VG00}. This approach unifies three seemingly distinct definitions of self-similarity that existed in the literature \cite{BWW00,Pec95,GW89,NTG97}. Importantly, each self-similar tree distribution corresponds to a unique Tokunaga sequence $T_k$ that quantifies merging of branches of distinct orders (Theorem \ref{thm:TokSeq}, Corollary \ref{cor:PruneVsToeplitz}). This provides a fundamental connection between an abstract measure-theoretic prune-invariance property and the tangible Tokunaga coefficients that can be statistically estimated in a single tree. 

The family of self-similar distributions (Definition \ref{def:SST}) rigorously reproduces the key geomorphic scalings discovered and reconfirmed during the past 80 years for river basins and summarized by \cite{RIR01,MRR+96,GTD07,Turcotte_book,DR00}, with a close fit to the examined exponents (Table \ref{tab:cTok}, Fig. \ref{fig:Beaver_Horton}).  Interestingly, this fit is achieved within a one-parameter family of critical Tokunaga trees (Definition \ref{def:TokTree}, Eq. \eqref{eqn:Tok}). Although trees that satisfy Eq. \eqref{eqn:Tok} (and commonly referred to as Tokunaga trees) have been known for a long time \cite{Pec95,Tok78,VG00}, only very recently a rigorous understanding has been gained of the theoretical importance of this constraint within the general framework of branching processes \cite{KZ20survey,KZ16,KZ18}. In addition, neither the order distribution nor link lengths distribution (specified by $\pi_K$ and $\lambda_j$ of Eq. \eqref{eqn:sstRAM}) have been examined in Tokunaga trees. This justifies our reference to the critical Tokunaga tree as a novel model. The critical Tokunaga model provides a natural parametric extension of the critical binary Galton-Watson branching process (and includes it as a special case with $c = 2$), which proved to be an indispensable model in many areas and remains at the forefront of theoretical and applied research nearly 150 years after its discovery \cite{GW}. This hints at deep and not fully understood symmetries in the structure of river networks. The theory of random self-similar trees explains the mathematical origin of these symmetries and offers tools for future exploration.

The presented results might advance applied statistical analysis of river stream attributes, via mapping all quantities of interest to a single master parameter $c$ of Eq. \eqref{eqn:Tok}. Statistical estimation of this parameter can be designed more effectively than that for a range of distinct yet possibly related quantities (e.g. Horton exponents). This in turn facilitates global mapping of river network features and studying possible effects of hydroclimatic variables on landscape dissection. Corollary \ref{cor:cRbRmRaRsRl} shows that multiple Horton laws examined in this work hold with only two distinct Horton exponents: $R_B = R_M = R_A = 2c$ and $R_L = R_S = R_\Lambda = c$. This substantial reduction of observed quantities is well supported by data (Table \ref{tab:cTok}, Fig. \ref{fig:Beaver_Horton}) and might inform a range of modeling and theoretical efforts. 

The critical Tokunaga model presents an ultimately symmetric class of trees characterized by coordination, Horton prune-invariance, criticality, time-invariance, and identically distributed link lengths (and hence local contributing areas). Despite these multiple constraints, this class is surprisingly rich, extending from perfect binary trees ($c = 1$) to the famous Shreve's random topology model ($c = 2$) to the structures reminiscent of the observed river networks ($c \approx 2.3$) and beyond. While offering a convenient theoretical and modeling paradigm, the critical Tokunaga model is merely a subclass of a much broader family of self-similar trees that might better accommodate for various problem-specific data features. For instance, Fig. \ref{fig:d} suggests that the observed stream networks tend to cluster around the critical Tokunaga line $R_B = 2R_L$ in the ($R_B$,$R_L$) space. An applied study can use the self-similar theory to either focus on the symmetries of the critical Tokunaga family, or explore deviations from this stiff parameterization, both of which may have physical underpinnings. 

Multiple properties of the critical Tokunaga family are well justified by the empirical evidence. We already mentioned that the coordination means that the basin structure is determined by its Horton-Strahler order, and the Horton prune-invariance implies that the fundamental scaling laws remain the same after changing data resolution. Criticality ensures that the stream networks uniformly fill the space, instead of exploding (supercritical case) or rapidly fading off (subcritical case). The time-invariance (invariance of basin order frequencies at different distances to the outlet) might reflect a physical process of formation of a stream network from sources downstream, so that a link only ``knows'' the information about the upstream part of the basin, yet remains unaware of how far it is from the outlet. Deviations from this invariance might point out to anthropogenic changes in a basin by which various downstream alterations (dam construction, sediment aggradation, etc.) impose upstream changes that deviate from the natural organization of a left-alone erosional landscape. In the same vein, it would be interesting to find a hydrogeomorphological explanation for the joint distribution of the merging sub-basins \eqref{eqn:joint}. 

The self-similar family extends beyond the hydrological constraints, allowing one to study self-similar trees with edge lengths that depend on the position within the hierarchy, arbitrary fractal dimension ${\bf d} > 1$, and arbitrary Horton exponents $R_B > 2$ and $R_L > 1$. For instance, the RAM might be a suitable model for phylogenetic trees \cite{A01} or dendritic structures generated by Diffusion Limited Aggregation (DLA). We recall that the geometric form of the Tokunaga coefficients $T_k\propto c^k$ with $c \approx  2.72 \pm 0.22$ has been known in DLA for a long time \cite{NTG97,O92}. It is noteworthy that the independently estimated fractal dimension of DLA clusters, ${\bf d} = 1.7\pm 0.05$ \cite{DNS86}, coincides with the fractal dimension of a critical Tokunaga tree with $c = 2.72\pm 0.22$ according to our Eq. \eqref{eqn:FDlogs}: ${\bf d} = \log_c(2c) = 1.7\pm 0.05$.

We conclude by suggesting that the understanding of the hierarchical organization and scaling in convergent (tributary) river networks gained here can be extended to other geomorphological processes. Important examples include dynamic reorganization of landscapes and stream networks \cite{FGD10,SRF15,WMPGC14} and scaling of peak flows \cite{MGM06}. 
Our results can also be extended to study the divergent (distributary) networks of river deltas that are commonly represented by a directed acyclic graph \cite{TLEZGRF17} -- a next step in complexity after trees examined in this work. Quantifying the structure, self-similarity, and scaling of such graphs contributes to a still-missing unifying theory explaining how deltaic river networks self-organize to distribute water and sediment fluxes to the shoreline \cite{TLZF15}.

\begin{sidewaystable*}
\centering
\vspace{10cm}
 \caption{{\bf Scaling exponents in critical Tokunaga model.}
 Selected scaling exponents (1st column) in critical Tokunaga model
 expressed via the model parameter $c\ge 1$ (2nd column), 
 fractal dimension ${\bf d}$ (3rd column), and
 Hack's exponent ${\bf h}$ (4th column).
 Columns 5-7 show the values of the exponents in critical Tokunaga model for $c=2.3, 2.4, 2.5$.
 For comparison, column 8 shows the values estimated in the OCN model. 
 Columns 9 summarizes estimations in the observed river networks. 
 The agreement of the exponents of the critical Tokunaga model with $c=2.3$ (column 6) with those observed 
 from real basins is noted. 
 \label{tab:cTok}}
 \medskip
 \begin{tabular}{c|c|c|c|c|c|c|c|c}
 Exponent  &\multicolumn{3}{c|}{Expressed via} & \multicolumn{3}{c|}{Critical Tokunaga model} & OCN$^\dagger$ & Real basins$^\ddagger$ \\
  	& $c$  	& ${\bf d}$ 	& ${\bf h}$ & $c=2.0^{*}$ & $c=2.3$ & $c=2.5$ &  & \\
\hline
$R_B=R_M=R_A$ 	& $2c$ 	& $2^{{\bf d}/({\bf d}-1)}$ & $2^{1/(1-{\bf h})}$ & 4 & 4.6 & 5.0 & 4 & 4.1 -- 4.8\\
$R_S=R_L$ 		& $c$	& $2^{1/({\bf d}-1)}$		& $2^{{\bf h}/(1-{\bf h})}$ & 2 & 2.3 & 2.5 & 2 & 2.1 -- 2.7\\
${\bf d}=\displaystyle\frac{\log R_B}{\log R_L}$ & $\log_c{(2c)}$ & ${\bf d}$ & ${\bf h}^{-1}$ & 2 & 1.832 & 1.756 & 2 & 1.7 -- 2.0\\
${\bf h}=\displaystyle\frac{\log R_L}{\log R_B}$ & $\log_{2c}c$ & ${\bf d}^{-1}$ & ${\bf h}$ & 0.5 & 0.546 & 0.569 & 0.57 & 0.5 -- 0.6\\
$\beta_A$ & $\log_{2c}2$ & $1-{\bf d}^{-1}$ & $1-{\bf h}$ & 0.5 & 0.454 & 0.431 & 0.43 & 0.4 -- 0.5\\
$\beta_{\Lambda}$ & $\log_{c}2$ & ${\bf d}-1$ & ${\bf h}^{-1}-1$ & 1 & 0.832 & 0.756 & 0.8 & 0.65 -- 0.9\\
\hline
\multicolumn{9}{p{.8\textwidth}}{$^*$ The same as the critical binary Galton-Watson model with i.i.d. exponential edge lengths, or Shreve's random topology model conditioned on the number of sources.}\\
\multicolumn{9}{p{.8\textwidth}}{$^\dagger$ Average values estimated in simulated OCN basins. According to \cite{RIR01,Cetal98}.}\\
\multicolumn{9}{p{.8\textwidth}}{$^\ddagger$ According to \cite{RIR01,DBE94,Pec95,MRR+96,ZZF13,Rigon+96,TBR88,TBR89,DR00}.}
\end{tabular}
\end{sidewaystable*}

%=======================================================
\section{Appendices}\label{sec:appdx}

\subsection{Horton-Strahler orders and Horton pruning}\label{appdx:HSorder}%%%%%%%%%%
The importance of links and junctions in the basin hierarchy is measured by the Horton-Strahler order $K \geq 1$ \cite{Horton45,Strahler}. Each link and its upstream junction have the same order. The order assignment is done in a hierarchical fashion, from the sources downstream. Each source is assigned order $1$. When two links of the same order $K$ merge at a junction, the junction is assigned order $K+1$. When two links with different orders $K_1 > K_2$ merge at a junction, the largest order prevails and the junction is assigned order $K_1$. The connected sequence of links and their upstream junctions of the same order $K$ is called a {\it stream} ({\it branch}) of order $K$. We denote by $N_K = N_K[T]$ the number of streams of order $K$ in a finite tree $T$. The Horton-Strahler order ${\sf ord}(T)$ of a tree $T$ is the maximal order of its links (junctions, streams). The Horton-Strahler ordering is illustrated in Fig. \ref{fig:Beaver}.

The Horton-Strahler orders are closely related to the operation of Horton pruning $\cR$ that removes the source links from a basin. This relation has been first recognized by Melton \cite{Melton59} and proved valuable in rigorous statistical analyses of tree self-similarity \cite{Pec95,BWW00,KZ20survey,PG99,KZ16}. Formally, we consider the map $\cR: \cL \to \cL$ that removes the source links from a tree $T$. This may create nonbranching chains of links connected by degree $2$ junctions -- every such chain is merged into a single link. The Horton pruning $\cR$ reduces the order of each surviving stream, and hence the basin order, by $1$. Accordingly, the order of a tree is the minimal number of Horton prunings that completely eliminates it, as in Eq. \eqref{eqn:ordTdef}. We emphasize that the pruning cannot cut a stream in the middle -- it can only eliminate the entire stream after a finite number of iterations \cite{Melton59}. Figure \ref{fig:Beaver} illustrates the Horton pruning for the stream network of Beaver Creek, KY -- the order of this basin is $K=6$ because it is eliminated in six Horton prunings.

\subsection{Asymptotic behavior of a sequence: Generating function approach}\label{appdx:asymp}%%%%%%%%%%%%
This section summarizes the basic facts about generating functions that are the main tool in establishing 
asymptotic behavior of stream attributes in a self-similar basin.
 
The {\it generating function} $f(z)$ of a sequence $a_k \ge 0$, $k=0,1,\dots$, of 
non-negative real numbers is defined as a {\it formal power series} 
\begin{equation}
f(z) = \sum_{k=0}^{\infty}a_kz^k,\quad z\in\mathbb{C}.
\label{GF}
\end{equation}
It is known \cite{Ahlfors} that there exist such a real number $r\ge 0$ that the series 
in the right hand side of \eqref{GF} converges to the function $f(z)$ for any 
$|z|<r$ and diverges for any $|z|>r$.
The number $r$ is called the {\it radius of convergence} 
of the sequence $a_k$;
it provides notable constraints on the asymptotic behavior of $a_k$.
The smaller the radius of convergence, the faster the growth of 
the sequence coefficients.
Informally, $0<r<1$ implies that the coefficients $a_k$ increase geometrically, $r>1$ that the coefficients decrease geometrically, and $r=1$ that the coefficient vary at a rate slower than geometric (e.g., polynomially). 
The values $r=0$ and $r=\infty$ imply a faster than geometric growth  
or decay, respectively.

The Cauchy-Hadamar theorem \cite{Ahlfors} expresses the radius of convergence in terms of the series coefficients:
\begin{equation}
\label{eq:CH}
\frac{1}{r} = \limsup_{k\to\infty} a_k^{1/k}.
\end{equation}
Often, the radius of convergence for $a_k$ can be easily found from the 
explicit form of $f(z)$.
Specifically, if $r>0$, then the function $f(z)$ is analytic within the
disk $|z|<r$ and has at least one {\it singularity} on the circle $|z|=r$,
that is it has to diverge for at least one point on that circle 
\cite[Thm.~2.4.2]{Wilf}.
Thus, the radius of convergence equals to the modulus of a singularity closest 
to the origin.
Furthermore, recalling that $a_k\ge 0$ we have
\begin{equation}
|f(z)|=\left|\sum_{k=0}^{\infty}a_k\,z^k\right| 
\le \sum_{k=1}^{\infty} a_k\,|z|^k=f(|z|),
\label{positive}
\end{equation}
where the equality is only achieved for $z=|z|$.
This means that the singularity closest to the origin lies on the real axis
(although there might be other singularities with the same modulus.)
This makes the search for such a singularity much easier: one can only consider
the restriction of the function $f(z)$ on the real axis.
In other words, despite the use of complex analysis in establishing 
some of our results, the applied treatment of suitable generating
functions can be done in the real domain.
Furthermore, if the singularity of $f(z)$ nearest to the origin is a simple pole,
then the coefficients $a_k$ asymptotically form a geometric series, which
we refer to as {\it Horton law}.

\begin{prop}[{\bf Horton Law for a Simple Pole Sequence}]
\label{P1}
Suppose $f(z)=\sum_{i=1}^{\infty}a_kz^k$ is analytic in the disk $|z|<\rho$
except for a single pole of multiplicity one at a positive real value $r<\rho$. 
Then the sequence $a_k$ obeys Horton law 
\begin{equation}
\label{eq:S}
\lim_{k\to\infty} a_k\,r^k = \alpha
\end{equation}
for some $0<\alpha<\infty$. 
Furthermore, if we define $g(z)=f(z)(z-r)$, then $\alpha=-g(r)/r$.
\end{prop}
\begin{proof}
\noindent We have, for any $\Delta \in (0,r)$ \cite{Ahlfors}:
\begin{equation}\label{eq:mk}
a_k = \frac{1}{2\pi i} \oint_{|z|=\Delta} \frac{f(z)dz}{z^{k+1}}.
\end{equation}
By the Residue Theorem \cite{Ahlfors}, we obtain, for any $\gamma \in (r,\rho)$
\begin{align}
\frac{1}{2\pi i} \oint_{|z|=\gamma} \frac{f(z)dz}{z^{k+1}}
& ={\sf Res}\left(\frac{f(z)}{z^{k+1}};0\right)
+{\sf Res}\left(\frac{f(z)}{z^{k+1}};r\right)\\
&= a_k+{\sf Res}\left(\frac{f(z)}{z^{k+1}};r\right).
\end{align}
Therefore,
\begin{equation}
	a_k= \frac{1}{2\pi i} \oint_{|z|=\gamma} \frac{f(z)dz}{z^{k+1}}
	-{\sf Res}\left(\frac{f(z)}{z^{k+1}};r\right),
\end{equation}
where
\begin{equation}
\left| \oint_{|z|=\gamma} \frac{f(z)dz}{z^{k+1}} \right|
\leq \frac{\max_{|z|=\gamma}|f(z)|}{\gamma^k}
= o\left(r^{-k}\right).
\end{equation}

\noindent Consider $g(z)=(z-r)f(z)$. 
It is known that \cite{Ahlfors}
\begin{equation}
{\sf Res}\left(f(z);r\right) = g(r),
\end{equation}
and hence
\begin{equation}
{\sf Res}\left(\frac{f(z)}{z^{k+1}};r\right) = \frac{g(r)}{r^{k+1}}
=\frac{g(r)}{r}r^{-k}.
\end{equation}

\noindent Accordingly, we obtain
\begin{equation}
	a_k = -\frac{g(r)}{r}r^{-k} +o(r^{-k}),
\end{equation}
which completes the proof.
\end{proof}

Proposition~\ref{P1} is used in Sect.~\ref{sec:HL},\ref{sec:HA} to establish 
Horton laws for $\Lambda_k$ and $A_k$.

\subsection{Power law distribution of link attributes}\label{appdx:linkatrib}%%%%%%%%%%
Consider the value $Z_{(i)}$ of an attribute $Z$ calculated at link $i$ in a large basin. The average number of links of order $K$ is given by $N_K S_K$, where $S_K$ denotes the average number of links within a stream of order $K$. One can heuristically approximate the frequencies of $\{Z_{(i)}\}$ by using the same average value $Z_K$ for any link of order $K$. Then, assuming Horton laws for $N_K$ and $S_K$ with exponents $R_B$ and $R_S$, respectively, and considering the limit of an infinitely large basin we find
\be
\#\{i:Z_{(i)}\ge R_Z^K\}\approx \sum_{j=K}^\infty N_j S_j \propto 
\!\!\sum_{j=K}^\infty \!\left(\frac{R_S}{R_B}\right)^j\propto \!\left(\frac{R_S}{R_B}\right)^K \!\!\!.
\ee
This is a punctured (by discrete order) version of a general power law relation of Eq. \eqref{eqn:HL2}.

\subsection{Self-similar trees, Tokunaga coefficients, Horton laws}
\label{appdx:sstTokHL}%%%%%%%%%%
\begin{Def}[Tokunaga coefficients] 
Fix a coordinated measure $\mu$ on $\cT$. For any pair $i < j$, the Tokunaga coefficient $T_{i,j} = T_{i,j}(\mu)$ is the expected number of streams of order $i$ per a randomly selected stream of order $j$ with respect to $\mu$ \cite{BWW00,Pec95,KZ20survey,DR00,Tok78}.
\end{Def}
We can arrange the Tokunaga coefficients for trees of a given order $K$ in an upper triangular matrix
\begin{equation}\label{tokuind1}
\mathbb{T}_K=\left[\begin{array}{ccccc}
0 & T_{1,2} & T_{1,3} & \hdots & T_{1,K} \\
0 & 0 & T_{2,3} & \hdots & T_{2,K} \\
0 & 0 & \ddots & \ddots & \vdots \\
\vdots & \vdots & \ddots & 0 & T_{K-1,K} \\
0 & 0 & \dots & 0 & 0\end{array}\right].
\end{equation}

\begin{thm}[Tokunaga sequence \cite{KZ20survey}]\label{thm:TokSeq} 
Suppose $\mu$ is a self-similar measure on $\cT$. Then the Tokunaga coefficients satisfy the Toeplitz property: $T_{i,i+k} = T_k$ for any positive integer pair $i$, $k$. In this case the Tokunaga matrix becomes Toeplitz:
\begin{equation}\label{tokuind3}
\mathbb{T}_K=\left[\begin{array}{ccccc}
0 & T_1 & T_2 & \hdots & T_{K-1} \\
0 & 0 & T_1 & \hdots & T_{K-2} \\
0 & 0 & \ddots & \ddots & \vdots \\
\vdots & \vdots & \ddots & 0 & T_1 \\
0 & 0 & \dots & 0 & 0\end{array}\right].	
\end{equation}
\end{thm}
\begin{proof}
Consider the pushforward measure $\cR_*(\mu)$ induced on $\cT$ by the Horton pruning operator:
\be
\cR_*(\mu)(A) = \mu\big(\cR^{-1}(A)\big),  \quad \forall A\subset \cT.
\ee
Since Horton pruning decreases the order of every stream by $1$, the Tokunaga coefficients $T_{i,j}^R$ computed on $\cT$ with respect to $\cR_*(\mu)$ satisfy $T_{i,j}^R = T_{i+1,j+1}$. The self-similarity of $\mu$ implies $T_{i,j}^R = T_{i,j}$. Combining these relations, we find $T_{i,j}  = T_{i+1,j+1}$. This establishes the desired Toeplitz property of the Tokunaga coefficients.
\end{proof}

We refer to the elements $T_k$ of the Tokunaga sequence as {\it Tokunaga coefficients}, which creates no confusion with the original double-indexed coefficients $T_{i,j}$.

\begin{cor}[Prune-invariance vs. Toeplitz]\label{cor:PruneVsToeplitz}
Suppose $\mu$ is a coordinated measure on $\cT$. Then the Toeplitz property $T_{i,i+k} = T_k$ and Horton prune-invariance of Eq. \eqref{eqn:ssm} are equivalent (i.e., both hold or do not hold at the same time).
\end{cor}
\begin{proof}
It has been shown in the proof of Theorem \ref{thm:TokSeq} that in coordinated trees both prune-invariance and Toeplitz property take the same algebraic form  $T_{i,j}  = T_{i+1,j+1}$.
\end{proof}

Consider the mean number $$\cN_i[K] ={\sf E}[N_i(T) \,|\,{\sf ord}(T) = K]$$ of streams of order $i$ in a basin of order $K$, and the mean magnitude (number of upstream sources) $M_i$ of a stream of order $i$. Observe that for a fixed $K$ the stream counts $\cN_i[K]$ form a decreasing sequence in $i$, and the sequence's first term $\cN_1[K]$ increases with $K$. At the same time, the average magnitudes $M_i$ form an increasing sequence in $i$ whose first terms are independent of basin order $K$ for any $K\geq i$. This explains the dependence on $K$ in the average stream counts and absence of such in the average magnitudes. The definition implies $\cN_K[K] = M_1 = 1$ and $\cN_1[K] = M_K$ for any tree distribution. Moreover, in self-similar trees the two sequences are deterministically related as \cite{Pec95,KZ16}
\be\label{eqn:N2M}
\cN_{K-j+1}[K]=\cN_1[j]=M_k \quad \text{for any}\quad 1\le j\le K.
\ee 

%\medskip

\subsection{Proof of Theorem \ref{thm:M}}\label{appdx:proof1}%%%%%%%%%%%
The average magnitude $M_k$ is the mean number of sources upstream of an order $k$ stream. 
It can be represented as the sum of the magnitudes of two order $k-1$ streams that formed this stream, plus the 
magnitudes of all its side tributaries. Hence $M_1=1$, and	
\begin{equation}\label{eqn:Mk}
M_k = 2\,M_{k-1}+\sum_{i=1}^{k-1}M_i\,T_{k-i},\quad \text{ for }\quad k>1.
\end{equation}
The generating function for the average magnitudes $M_k$ is obtained by multiplying both 
sides in \eqref{eqn:Mk} by $z^k$ and summing over $k\geq 1$:
\[M(z) =\sum_{k=1}^\infty M_k\, z^k =z+2zM(z)+M(z)\,T(z).\]
Thus,	
\begin{equation}\label{eqn:M}
M(z) = \frac{z}{1-2z-T(z)}=-{z \over t(z)},
\end{equation}
where, according to Eq. (16) of the main text:
\[t(z)=-1+2z+T(z),\quad T(z) = \sum_{i=1}^{\infty}T_kz^k.\]
The function $M(z)$ is analytic with the exception of zeroes and singularities of $t(z)$. 
Observe that $t(0)=-1$, and since $T_k\ge 0$ we have $t(1/2)=T(1/2)\ge 0$.
Furthermore, since
\[{d \over dz}t(z)=2+\sum_{k=1}^\infty \,kT_k\,z^{k-1} >0\,\quad\text{for all } z\in(0,\infty),\] 
the equation $t(z)=0$ has a unique real root $w_0$ of 
multiplicity one in the interval $(0,1/2]$.
Let $r_T$ be the radius of convergence for $T(z)$, and hence for $t(z)$.
We notice that $r_T>w_0$, so the radius of convergence of $M(z)$ coincides
with the root of $t(z)$ closest to the origin.  
We claim that this root is $w_0$.  
Assume otherwise, so there exists $w\in\mathbb{C}$ such that $t(w)=0$ and $|w|<w_0$. 
Since $w_0$ is the unique real root of $t(z)$ within $(0,1/2]$, $w$ must have a non-zero
imaginary part. 
This means that the singulatiry of $M(z)$ closest to the origin is not on the real
axis, which contradicts \eqref{positive}.
Hence the radius of convergence of $M(z)$ is $w_0$, and $w_0$ is a simple pole of $M(z)$.
Proposition~\ref{P1} now establishes the result. 
\qed

\subsection{Exact Horton law}\label{appdx:eHorton}%%%%%%%%%%%%
Assume that the Horton law for stream numbers $\mathcal{N}_1[K]$, and hence for magnitudes $M_K$,
holds exactly (recall that $M_1=1$):
\begin{equation}
\label{eq:Hexact}
M_K = R_M^{K-1}.
\end{equation}
Then, 
\[M(z) = \frac{z}{1-R_Mz}\] 
which leads to
\[t(z) = -\frac{z}{M(z)}=-1+R_Mz\quad\text{and}\quad T(z) = (R_M-2)z.\]
This implies that the only self-similar model with exact Horton law 
corresponds to the Tokunaga sequence
\[T_1 = R_M - 2, \quad T_k = 0\quad\text{for}\quad k>1.\]

\subsection{Scalings in a critical Tokunaga tree}
\label{appdx:cTok}%%%%%%%%%%

The critical Tokunaga model (Definition \ref{def:TokTree}) has the following exact form of the average branch counts and average magnitudes (Cor. 4 in \cite{KZ20survey}):
\be
\cN_{K-i+1}[K]=M_i={R_B^i+R_B-2 \over 2(R_B-1)}  \quad \text{for any}~~ 1\le i\le K.
\ee
In addition, it satisfies the Horton law for the original stream counts $N_i$ 
\cite[Cor.~5]{KZ20survey}:
\be
{N_i \over N_{i+1}} \stackrel{p}{\longrightarrow} R_B \quad \text{ as }\, i \to \infty,
\ee
where $\stackrel{p}{\longrightarrow}$ denotes convergence in probability \cite{BW07}. This result strengthens the statement of Theorem \ref{thm:M}, Eq. \eqref{eqn:meanHorton} that is formulated for the respective averages. Finally, the weak law of large numbers holds for the tree order. Formally, denote by $T[K]$ a critical Tokunaga tree of order $K$ and write $\#T[K]$ for the number of links in this tree. Then  \cite[Cor.~6]{KZ20survey}
\be
{\log_{R_B}\#T[K] \over K} \stackrel{p}{\longrightarrow} 1 \quad \text{ as }\, K \to \infty.
\ee
Informally, this means that the tree order grows as a logarithm base $R_B$ of the tree size.

The identically distributed link lengths imply identically distributed local areas, which in turn establishes the Horton law for $A_i$. Specifically, in a critical Tokunaga tree we have (Appendix \ref{sec:HA}):
\begin{equation}\label{eqn:AvsM}
A_i \sim {\rm Const.} \times M_i,
\end{equation}
where $x_i\sim y_i$ stands for $\displaystyle\lim_{i\to\infty}x_i/y_i = 1$.
The same approach shows that the asymptotic of Eq. \eqref{eqn:AvsM} holds also for the average total channel length $L^{tot}_i$ upstream of a stream of order $i$, with different proportionality constant. The asymptotic of Eq. \eqref{eqn:AvsM} formalizes one of the key empirical observations \cite{RIR01} that connects a physical (area $A_i$) and a combinatorial (magnitude $M_i$) attributes of a river basin. This asymptotic may not hold in a general self-similar (not critical Tokunaga) tree.

\subsection{Proofs of Corollary \ref{cor:HBPL}, Theorem \ref{thm:HacksLawTok}, Corollary \ref{cor:cRbRmRaRsRl}}\label{appdx:ctcProof}%%%%%%%%%%
\begin{proof}[Proof of Corollary \ref{cor:HBPL}]
By definition of RAM, the length of a stream of order $j$ is an exponential random variable with rate $\lambda_j$. 
In a self-similar tree, the rate is given by Eq. \eqref{eqn:Tok}: $\lambda_j = \gamma\, \zeta^{1-j}$. This implies $L_j = \lambda_j^{-1} = \gamma^{-1}\, \zeta^{j-1}$, 
which is equivalent to the statement of Theorem \ref{thm:HacksLawTok} (Eq. \eqref{eqn:HacksLawTok}).
\end{proof}

\begin{proof}[Proof of Theorem \ref{thm:HacksLawTok}]
Recall the Horton law for the average magnitude (Theorem \ref{thm:M}, Eq. \eqref{eqn:MNk}) that holds in any tree with a tamed Tokunaga sequence \mbox{($\limsup\limits_{k \to \infty} T_k^{1/k}<\infty$)} and the Horton law for the average length of the longest stream (Theorem \ref{thm:HortonLawLambda}, Eq.  \eqref{eqn:Hheight}) that holds in any self-similar RAM tree. These laws apply to a critical Tokunaga tree of the current statement.  Furthermore, the asymptotic equivalence between the average basin contributing area and average basin magnitude (Eq. \eqref{eqn:AvsM}) implies the Horton law for the average basin areas with Horton exponent $R_M$. Finally, we use the general result of Eq. \eqref{eqn:hRlRa} to establish the Hack's law (Eq. \eqref{eqn:HacksLawTok}) in a critical Tokunaga tree.
\end{proof}

\begin{proof}[Proof of Corollary \ref{cor:cRbRmRaRsRl}]
Using the definition of $\hat{t}(z)$ (Eq. \eqref{eqn:tofz}) and the geometric form of the Tokunaga coefficients (Eq. \eqref{eqn:Tok}) we obtain $\hat{t}(z) = (1 - 2cz)(z - 1)/(1 - cz)$. The only real root of $\hat{t}(z)$ within $(0,1/2]$ is $w_0 = (2c)^{-1}$. By Theorem \ref{thm:M}, Eq. \eqref{eqn:RbRmW0} we have $R_B = R_M = 2c$, and Eq. \eqref{eqn:Ck} implies $S_K = c^{K-1}$, which corresponds to $R_S = c$. The equality $R_S = R_L$ follows from independence of the distribution of link lengths of their position within a basin. Finally, $R_\Lambda = c$ is established in Theorem \ref{thm:HortonLawLambda}, Eq.  \eqref{eqn:Hheight}. 
\end{proof}

\subsection{Fractal dimension of a self-similar RAM tree}\label{appdx:fracD}%%%%%%%%%%%%%%
Consider a self-similar RAM tree $T$ (Theorem 3) with a Tokunaga sequence $\{T_k\}$ satisfying 
$\limsup_{k \to \infty} T_k^{1/k} < \infty$, and parameters $\gamma>0$ and $\zeta >1$. 
Below we construct a Markov tree process $\{\Upsilon_K\}_{K=1,2,\hdots}$ 
corresponding to $T$ following \cite{KZ20survey} and use it to 
find the fractal dimension of the resulting tree in the limit of
infinite tree order.
The construction below closely reproduces that of the RAM (see the main text), but 
scales the edge lengths so that an infinitely large tree has a proper fractal dimension.

Let $\Upsilon_1$ be an I-shaped tree of Horton-Strahler order one, with the edge length distributed as an 
exponential random variable with parameter $\gamma$.
Conditioned on $\Upsilon_K$, the tree $\Upsilon_{K+1}$ is constructed according to the following transition rules.    
We attach new leaf edges to $\Upsilon_K$ at the points  sampled by an inhomogeneous Poisson point process 
with the intensity $\rho_{j,K}=\gamma \zeta^{K-j}T_j$ along the edges of order $j\leq K$ in $\Upsilon_K$. 
We also attach a pair of new leaf edges to each of the leaves in $\Upsilon_K$.
The lengths of all the newly attached leaf edges are i.i.d. exponential random variables with parameter $\gamma \zeta^K$
that are independent of the combinatorial shape and the edge lengths in $\Upsilon_K$. 
Finally, we let the tree $\Upsilon_{K+1}$ consist of $\Upsilon_K$ and all the attached leaves and leaf edges.

By construction, a branch of order $j$ in $\Upsilon_K$ becomes a branch of order $j+1$ in $\Upsilon_{K+1}$ after 
the attachment of new leave edges. 
The length of order $j$ branch in $\Upsilon_K$ (and therefore, the length of order $j+1$ branch in $\Upsilon_{K+1}$) 
is exponential random variable with parameter $\gamma \zeta^{K-j}$. 
Therefore, in a tree $\Upsilon_{K+1}$, the number $n_{1,j+1}(K+1)$ of side-branches of order one in a branch of 
order $j+1$ has geometric distribution:
\begin{align}
{\sf P}\big(n_{1,j+1}(K+1)=r \big)&={\gamma \zeta^{K-j} \over \gamma \zeta^{K-j}+\rho_{j,K}} \left(\!{\rho_{j,K} \over \gamma \zeta^{K-j}+\rho_{j,K}}\!\right)^r\nonumber\\
&=\frac{1}{1+T_j}\left(\frac{T_j}{1+T_j}\right)^r 
\end{align}
for $r=0,1,2,\hdots$, with the mean value
\[{\sf E}\big[n_{1,j+1}(K+1) \big]={\rho_{j,K} \over \gamma \zeta^{K-j}}=T_j.\]
After $i \geq 1$ rounds of attachments the mean number $n_{i,j+i}(M)$ of side-branches of order $i$ in a branch of 
order $j+i$ in a tree $\Upsilon_M$ (where $M=K+i$ and $K\geq j$) is 
$${\sf E}\big[n_{i,j+i}(M) \big]=T_j.$$
Each tree $\Upsilon_K$ is distributed as a self-similar RAM tree \cite{KZ20survey}
with Tokunaga sequence $\{T_k\}$ and parameters $(\gamma,\,\zeta)$, conditioned on 
its Horton-Strahler order being equal to $K$, and with its edge lengths scaled by $\zeta^{1-K}$.

Observe that by construction $\Upsilon_K \subset \Upsilon_{K+1}$.
Accordingly, there exists the limit space 
\[\Upsilon_\infty=\lim\limits_{K \rightarrow \infty} \Upsilon_K =\bigcup\limits_{K=1}^\infty \Upsilon_K.\]
The self-similarity of the RAM process suggests that the limit space does not change its statistical properties 
after rescaling, which corresponds here to the Horton pruning.
Let ${\bf d}$ denote its fractal dimension. 
That the limit space includes at least the root branch $\Upsilon_1$ implies ${\bf d}\ge 1$.
Assume that ${\bf d>1}$. 
Then, denoting the mean ${\bf d}$-dimensional volume of $\Upsilon_\infty$ by ${\bf vol}$, we have
\begin{equation}
\label{eqn:FDvol}
{\bf vol}=\sum\limits_{k=1}^\infty t_k {{\bf vol} \over \zeta^{{\bf d}k}}.
\end{equation}
This equation is obtained by splitting a tree $\Upsilon_\infty$ into the subtrees attached to its highest-order branch $\Upsilon_1$. 
There is an average of $t_1=T_1+2$ subtrees distributed as $\Upsilon_\infty$ scaled by $\zeta^{-1}$. 
In general, for each $k$, there will be an average of $t_k$ subtrees distributed as $\Upsilon_\infty$ scaled by $\zeta^{-k}$.
Scaling the lengths by $\zeta^{-k}$ in the ${\bf d}$-dimensional space results in scaling the volume by $\zeta^{-{\bf d}k}$.
The ${\bf vol}$ term in \eqref{eqn:FDvol} can be cancelled out, yielding
\begin{equation}
\label{eqn:FDthat}
\hat{t}\big(\zeta^{-{\bf d}}\big)=0,
\end{equation}
and hence, $\zeta^{-{\bf d}}=w_0=R_B^{-1}$. 
This leads to
\eqref{eqn:FDlogs}.

\subsection{Horton law for $\Lambda_k$}\label{sec:HL}%%%%%%%%%%%%%
If $T$ is the tree representing a stream network, then the length of the longest stream is 
the height of the tree $T$, denoted by $\textsc{height}(T)$ \cite{Pitman,KZ20survey}.

Consider a tree $T$ generated by a self-similar RAM
with a Tokunaga sequence $\{T_k\}$ satisfying $\limsup\limits_{k \to \infty} T_k^{1/k} < \infty$, 
and parameters $\gamma>0$ and $\zeta>1$. 
Let
\begin{equation}
\Lambda_k={\sf E}\left[\textsc{height}(T) \, \Big| \, {\sf ord}(T)=k\right]
\end{equation}
that represents the mean length of {\it  the longest river stream} 
in a basin with the Horton-Strahler order $k$.
Notice that, since $\textsc{height}\big(\mathcal{R}(T)\big) \leq \textsc{height}(T)$,
\begin{align}\label{eqn:ExpTreeHeightL}
\zeta \,\Lambda_{k-1}&={\sf E}\left[\textsc{height}\big(\mathcal{R}(T)\big)\, \Big| \, {\sf ord}(T)=k\right] \nonumber \\
&\leq {\sf E}\left[\textsc{height}(T)\, \Big| \, {\sf ord}(T)=k\right]=\Lambda_k.
\end{align}
Hence, since $\Lambda_1=\gamma^{-1}$, we have $\Lambda_k\geq\gamma^{-1}\zeta^{k-1}$. 
Next, let 
\[Y_1,Y_2,\hdots,Y_{N_1[T]}\]
denote the leaf lengths in the tree $T$. Then, since
\[\textsc{height}(T) \leq \textsc{height}\big(\mathcal{R}(T)\big) +\max\limits_{j=1,\hdots,N_1[T]}Y_j,\]
we have, 
\begin{align}\label{eqn:ExpTreeHeightR}
\Lambda_k &\leq {\sf E}\left[\textsc{height}\big(\mathcal{R}(T)\big)\, \Big| \, {\sf ord}(T)=k\right]\nonumber \\
&\qquad \quad +{\sf E}\left[\max\limits_{j=1,\hdots,N_1[T]}Y_j \, \Big| \, {\sf ord}(T)=k\right] \nonumber \\
&=\zeta\,\Lambda_{k-1} \,+ \gamma^{-1}{\sf E}\left[\sum\limits_{j=1}^{N_1[T]}{1 \over j}\, \Big| \, {\sf ord}(T)=k\right]\nonumber \\
&\leq \zeta\,\Lambda_{k-1} \,+ \gamma^{-1}{\sf E}\left[1+\log\big(N_1[T]\big)\, \Big| \, {\sf ord}(T)=k\right] \nonumber \\
&\leq \zeta\,\Lambda_{k-1} \,+ \gamma^{-1}+\gamma^{-1}\log\left({\sf E}\big[N_1[T]\, \big| {\sf ord}(T)=k\big]\right)
\end{align}
by Wald's equation, the Coupon Collector Problem, and finally, the Jensen's inequality.
Recall (Theorem \ref{thm:M}) the Horton law for the leaf count in a 
self-similar process
\[\mathcal{N}_1[k] = M_k =  M\,R_B^k+o\left(R_B^k\right).\]
Hence, equations \eqref{eqn:ExpTreeHeightL} and \eqref{eqn:ExpTreeHeightR} imply
\[0 ~\leq ~\Lambda_k-\zeta\,\Lambda_{k-1} ~\leq~\gamma^{-1}k\log{R_B}+\beta\]
for some constant $\beta$, and
\begin{equation}
\label{eqn:LkPowerLaw0}
0 ~\leq ~{\Lambda_k \over \Lambda_{k-1}}-\zeta ~\leq~\gamma^{-1}{k\log{R_B}+\beta \over \Lambda_{k-1}} \leq {k\log{R_B}+\beta \over \zeta^{k-2}} \to 0
\end{equation}
as $\,k\to\infty$. Accordingly,
\begin{align}\label{eqn:LkPowerLaw}
\log{\Lambda_k}&=\sum\limits_{j=2}^k \log\left({\Lambda_k \over \Lambda_{k-1}}\right)+\log{\Lambda_1} \nonumber \\
&=(k-1)\log\zeta+\sum\limits_{j=2}^k \log(1+\mathcal{E}_j) -\log\gamma,
\end{align}
where $0 \leq \mathcal{E}_j \leq {(k\log{R_B}+\beta)\zeta^{1-k}}$, and therefore, 
$\sum\limits_{j=2}^\infty \log(1+\mathcal{E}_j)$ converges to a constant.
We therefore conclude that the strong Horton law holds for $\Lambda_k$ with
Horton exponent $R_{\Lambda} = R_L = \zeta$:
\begin{equation}
\label{eq:Hheight}
\Lambda_k \sim \mathrm{Const.}\times \zeta^k.
\end{equation}

\subsection{Horton law for $A_k$}\label{sec:HA}%%%%%%%%%
Assume that the mean local contributing area of a link of order $k$ equals $\alpha_k$. 
Then the total mean contributing area $A_K$ of a tree of order $K \geq 1$ is
\begin{equation}\label{eqn:Ak0}
A_K=\sum\limits_{i=1}^K \alpha_i S_i \mathcal{N}_i[K],
\end{equation} 
where $S_k \mathcal{N}_k[K]$ is the mean number of links of order $k$ in a tree of order $K$.
A convenient recursive expression is obtained by noticing that $A_1=\alpha_1$ and
\begin{equation}\label{eqn:Ak}
A_K %= 2A_{k-1}+\alpha_k \left(1+\sum_{i=1}^{k-1}T_i \right)+\sum_{i=1}^{k-1}A_i\,T_{k-i}
= 2A_{K-1}+\alpha_K S_K+\sum_{i=1}^{K-1}A_i\,T_{K-i} \quad \text{ for }\quad K\geq 2.
\end{equation}		
The generating function for $A_k$ is given by
\[A(z)=\sum\limits_{k=1}^\infty A_k z^k=2zA(z)+\sum\limits_{k=1}^\infty \alpha_k S_k z^k +A(z)T(z),\]
which yields
\begin{equation}\label{eqn:A}
A(z)={\sum\limits_{k=1}^\infty \alpha_k S_k z^k  \over 1-2z-T(z)}
= -\frac{D(z)}{t(z)} = M(z)\frac{D(z)}{z}.
\end{equation}
Here $D(z)$ is the generating function for the mean local contributing areas 
$\alpha_k\,S_k$ of streams of order $k$. 
Suppose that the radius of convergence of $D(z)$ is larger than $w_0$.
Then, by Prop.~\ref{P1}, 
\begin{equation}
\label{eq:AM}
A_k \sim {\rm Const.}\times w_0^{-k}\sim {\rm Const.}\times M_k.
\end{equation}

Consider the critical Tokunaga model. Here $\alpha_k=\alpha$, $S_k = c^{k-1}$
and hence
\[D(z) = \frac{\alpha\,z}{1-cz}\]
whose radius of convergence $c^{-1}$ coincides with that of $t(z)$. 
Observe that the radius of convergence of $t(z)$ must be greater than its zero, hence $w_0<c^{-1}$,
and so the asymptotic of \eqref{eq:AM} holds.

%=======================================================

%\bibliography{bib_TokunagaTrees.bib.tex}

\begin{thebibliography}{99}
\bibitem{Horton45} 
R.\,E.~Horton,  
\newblock Erosional development of streams and their drainage basins; hydrophysical approach to quantitative morphology. 
\newblock Bulletin of Geophysical Society of America, \textit{56}, 275--370 (1945).

\bibitem{RIR01}
I.~Rodriguez-Iturbe and A.~Rinaldo, 
\newblock Fractal river basins: chance and self-organization. 
\newblock Cambridge University Press  (2001).

\bibitem{Kirchner93}
J.\,W.~Kirchner, 
\newblock Statistical inevitability of Horton's laws and the apparent randomness of stream channel networks. 
\newblock Geology, {\it 21}(7) 591--594  (1993).

\bibitem{Shreve66}
R.\,L.~Shreve , 
\newblock Statistical law of stream numbers. 
\newblock J. Geol., \textit{74}(1) 17--37 (1966).

\bibitem{BWW00}
\newblock  G.~Burd,  E.\,C. Waymire, and R.\,D.~Winn, 
\newblock A self-similar invariance of critical binary Galton-Watson trees. 
\newblock Bernoulli Society for Mathematical Statistics and Probability, \textit{6}(1), 1--21 (2000).

\bibitem{Pitman}
J.~Pitman, 
\newblock Combinatorial Stochastic Processes. 
\newblock Ecole d'\'{e}t\'{e} de probabilit{\'e}s de Saint-Flour XXXII-2002.
\newblock Lectures on Probability Theory and Statistics, Springer, (2006). 

\bibitem{DBE94}
H.~De Vries,  T.~Becker,  and  B. Eckhardt,  
\newblock Power law distribution of discharge in ideal networks.
\newblock Water Resources Research, 30(12), 3541--3543 (1994).

\bibitem{Pec95} 
S.\,D.~Peckham, 
\newblock New Results for Self-Similar Trees with Applications to River Networks. 
\newblock Water Resources Research, \textit{31}(1), 1023--1029  (1995).

\bibitem{Sch67}
A.\,E.~Scheidegger,  
\newblock A stochastic model for drainage patterns into an intramontane treinch. 
\newblock Hydrological Sciences Journal, 12(1), 15--20 (1967).

\bibitem{TNT88}
H.~Takayasu, I.~Nishikawa, and H.~Tasaki,  
\newblock Power-law mass distribution of aggregation systems with injection.
\newblock Physical Review A, 37(8), 3110  (1988).

\bibitem{RRR+92}
A.~Rinaldo, I.~Rodriguez-Iturbe, R.~Rigon, R.\,L.~Bras, E.~Ijjasz-Vasquez, and A.~Marani,  
\newblock Minimum energy and fractal structures of drainage networks. 
\newblock Water Resources Research, 28(9), 2183--2195 (1992).

\bibitem{RRR+93}
R.~Rigon, A.~Rinaldo, I.~Rodriguez-Iturbe, R.\,L.~Bras, and E.~Ijjasz-Vasquez, 
\newblock Optimal channel networks: a framework for the study of river basin morphology.
\newblock  Water Resources Research, 29(6), 1635--1646 (1993).

\bibitem{RRRIB93}
A.~Rinaldo, I.~Rodriguez-Iturbe, R.~Rigon, E.~Ijjasz-Vasquez, and R.\,L.~Bras,   
\newblock Self-organized fractal river networks. 
\newblock Physical Review Letters, 70(6), 822 (1993).

\bibitem{MRR+96}
A.~Maritan, A.~Rinaldo, R.~Rigon,  A.~Giacometti,and I.~Rodriguez-Iturbe,  
\newblock Scaling laws for river networks.
\newblock Physical Review E, 53(2), 1510 (1996).

\bibitem{RRBMR14}
A.~Rinaldo, R.~Rigon, J.\,R. Banavar, A. Maritan, I. Rodriguez-Iturbe,  
\newblock Evolution and selection of river networks: Statics, dynamics, and complexity. 
\newblock Proceedings of the National Academy of Sciences, 111(7), 2417--2424 (2014).

\bibitem{BBB+18}
P.~Balister, J.~Balogh,  E.~Bertuzzo, B.~Bollob\'as, G.~Caldarelli, A.~ Maritan, R.~Mastrandrea, R.~Morris, and A.~Rinaldo, 
\newblock River landscapes and optimal channel networks. 
\newblock Proceedings of the National Academy of Sciences, 115(26), 6548--6553  (2018).

\bibitem{GW89}
V.\,K.~Gupta and E.\,C.~Waymire, 
\newblock Statistical self-similarity in river networks parameterized by elevation.
\newblock Water Resources Research, {\it 25}(3) 463--476 (1989).

\bibitem{GTD07}
V.\,K.~Gupta, B.\,M.~Troutman, and D.\,R.~Dawdy,
\newblock Towards a nonlinear geophysical theory of floods in river networks: an overview of 20 years of progress. 
\newblock Nonlinear Dynamics in Geosciences (pp. 121--151), Springer, New York, NY   (2007).

\bibitem{Turcotte_book}
D.\,L.~Turcotte,  
\newblock Fractals and chaos in geology and geophysics.
\newblock Cambridge University Press, (1997).

\bibitem{NTG97} 
W.\,I.~Newman, D.\,L.~Turcotte and A. M.~Gabrielov, 
\newblock Fractal trees with side-branching. 
\newblock Fractals, \textit{5}, 603--614  (1997).

\bibitem{PG99}
S.\,D.~Peckham and V.\,K.~Gupta,
\newblock A reformulation of Horton's laws for large river networks in terms of statistical self-similarity. 
\newblock Water Resources Research, {\it 35}(9), 2763--2777  (1999).

\bibitem{F+79}
P. Flajolet, J.-C. Raoult, and J. Vuillemin, 
\newblock The number of registers required for evaluating arithmetic expressions.
\newblock Theoretical Computer Science 9(1) 99--125 (1979).

\bibitem{DP06}
M. Drmota and H. Prodinger, 
\newblock The register function for t-ary trees ACM.
\newblock Transactions on Algorithms 2 (3)  318--334 (2006).

\bibitem{BP04}
M.~Baiesi and M.~Paczuski,  
\newblock Scale-free networks of earthquakes and aftershocks. 
\newblock Physical Review E, 69(6), 066106 (2004).

\bibitem{HTR08}
J.\,R.~Holliday, D.\,L.~Turcotte, and J.\,B.~Rundle,  
\newblock Self-similar branching of aftershock sequences.
\newblock Physica A: Statistical Mechanics and its Applications,  387(4) 933--943 (2008).

\bibitem{Y13}
M.\,R. Yoder, J. Van Aalsburg, D. L. Turcotte, S. G. Abaimov, and J. B. Rundle,  
\newblock Statistical variability and Tokunaga branching of aftershock sequences utilizing BASS model simulations. 
\newblock Pure and Applied Geophysics, (2013) 170(1-2) 155--171.

\bibitem{ZBZ13}
I.~Zaliapin, and Y.~Ben-Zion,  
\newblock Earthquake clusters in southern California I: Identification and stability. 
\newblock Journal of Geophysical Research: Solid Earth, 118(6), 2847--2864 (2013).

\bibitem{Kassab00}
G. S. Kassab, 
\newblock The coronary vasculature and its reconstruction.
\newblock Annals of Biomedical Engineering,  28(8) 903--915 (2000).

\bibitem{Cetal06}
F. Cassot, F. Lauwers, C. Fouard, S. Prohaska, and V. Lauwers-Cances,  
\newblock A novel three-dimensional computer-assisted method for a quantitative study of microvascular networks of the human cerebral cortex.
\newblock Microcirculation,  13(1) 1--18 (2006).

\bibitem{CLF07}
E.\,H. Campbell Grant, W.\,H. Lowe, and W.\,F. Fagan,   
\newblock Living in the branches: population dynamics and ecological processes in dendritic networks.
\newblock Ecology Letters,  10(2) 165--175 (2007).

\bibitem{TPN98}
D.\,L.~Turcotte, J.\,D.~Pelletier,  and W.\,I.~Newman,
\newblock Networks with side-branching in biology. 
\newblock Journal of Theoretical Biology, {\it 193}(4), 577--592 (1998).

\bibitem{KZ20survey} 
Y.~Kovchegov and I.~Zaliapin, 
\newblock Random Self-Similar Trees: A mathematical theory of Horton laws. 
\newblock Probability Surveys, \textit{17}, 1--213  (2020).

\bibitem{Strahler}
A.\,N.~Strahler,  
\newblock Quantitative analysis of watershed geomorphology.
\newblock Trans. Am. Geophys. Un., \textit{38} 913--920 (1957).

\bibitem{Melton59} 
M.\,A.~Melton, 
\newblock A derivation of Strahler's channel-ordering system.
\newblock The Journal of Geology, 67(3), 345--346 (1959).

\bibitem{Leopold} 
L.~Leopold,  M.\,G.~Wolman, and J.~Miller, 
\newblock Fluvial processes in geomorphology.
\newblock Dover Publications, Inc., New York, USA  (1992).

\bibitem{Tarboton96}
D.\,G.~Tarboton,
\newblock Fractal river networks, Horton's laws and Tokunaga cyclicity. 
\newblock Journal of hydrology, {it 187}(1)  105--117  (1996).

\bibitem{GW98}
V.\,K.~Gupta and E.\,C.~Waymire,
\newblock Some mathematical aspects of rainfall, landforms and floods.
\newblock In {\it O.\,E.~Barndorff-Nielsen, V.\,K.~Gupta, V.~Perez-Abreu, E.\,C.~Waymire (eds) Rainfall, Landforms and Floods}, 
Singapore: World Scientific, Singapore (1998).

\bibitem{ZZF13} 
S.~Zanardo,  I.~Zaliapin, and E.~Foufoula-Georgiou, 
\newblock Are American rivers Tokunaga self-similar? (in New results on fluvial network topology and its climatic dependence). 
\newblock Journal of Geophysical Research: Earth Surface, \textit{118}, 1--18 (2013).

\bibitem{Mesa18}
O.\,J.~Mesa,  
\newblock Cuatro modelos de redes de drenaje. 
\newblock Revista de la Academia Colombiana de Ciencias Exactas, F\'{i}sicas y Naturales, {\it 42}(165), 379--391 (2018).

\bibitem{Hack57}
J.\,T.~Hack, 
\newblock Studies of longitudinal stream profiles in Virginia and Maryland.
\newblock US Government Printing Office, {\it 294} (1957).

\bibitem{Rigon+96}
R.~Rigon, I.~Rodriguez-Iturbe, A.~Maritan, A.~Giacometti, D.~Tarboton, and A.~Rinaldo,
\newblock On Hack's Law. 
\newblock Water Resources Research,  \textit{32}(11), 3367--3374 (1996).

\bibitem{TBR88}
D.\,G.~Tarboton,   R.\,L.~Bras,  and I.~Rodriguez-Iturbe, 
\newblock The fractal nature of river networks.
\newblock Water Resources Research, 24, 1317--1322  (1988).

\bibitem{RI92}
 I.~Rodriguez-Iturbe, E.\,J.~Ijjasz-Vasquez, R.\,L.~Bras, and D.\,G.~Tarboton, 
\newblock Power law distributions of discharge mass and energy in river basins. 
\newblock Water Resources Research, {\it 28}(4) 1089--1093 (1992).

\bibitem{KZ16} 
Y.~Kovchegov and I.~Zaliapin, 
\newblock Horton Law in Self-Similar Trees.
\newblock Fractals, \textit{24}, 1650017  (2016).

\bibitem{LR89}
P.~La Barbera and R.~Rosso,  
\newblock On the fractal dimension of stream networks. 
\newblock Water Resources Research, 25(4), 735--741 (1989).

\bibitem{TBR89}
D.\,G.~Tarboton,   R.\,L.~Bras,  and I.~Rodriguez-Iturbe, 
\newblock Scaling and elevation in river networks. 
\newblock Water Resources Research, 25(9), 2037--2051 (1989).

\bibitem{Tok78} 
E.~Tokunaga,
\newblock Consideration on the composition of drainage networks and their evolution. 
\newblock Geographical Reports of Tokyo Metropolitan University, \textit{13}, 1--27 (1978).

\bibitem{VG00}
S. A.~Veitzer and V.\,K.~Gupta,
\newblock Random self-similar river networks and derivations of generalized Horton Laws in terms of statistical simple scaling. 
\newblock Water Resour. Res., {\it 36}(4) 1033--1048 (2000). 

\bibitem{KZ18}
Y.~Kovchegov and I.~Zaliapin, 
\newblock Tokunaga self-similarity arises naturally from time invariance. 
\newblock Chaos: An Interdisciplinary Journal of Nonlinear Science, {\it 28}(4) 041102  (2018).

\bibitem{EVC18}
E.\,V.~Chunikhina,
\newblock Entropy rates for Horton self-similar trees.
\newblock Chaos, {\bf 28}(8), 081104 (2018).
%\url{https://doi.org/10.1063/1.5048965}

\bibitem{EVCthesis}
E.\,V.~Chunikhina,  
\newblock Information theoretical analysis of self-similar trees.
\newblock Ph.D. thesis, Oregon State University (2018).

\bibitem{DR00}
P.~Dodds and D.~Rothman,  
\newblock Scaling, Universality, and Geomorphology.
\newblock Ann. Rev. Earth and Planet. Sci., {\it 28} 571--610 (2000).

\bibitem{GCO96}
V.\,K.~Gupta, S.\,L.~Castro, and T.\,M.~Over, 
\newblock On scaling exponents of spatial peak flows from rainfall and river network geometry.
\newblock Journal of Hydrology, {\it 187}(1-2) 81--104  (1996).

\bibitem{Cetal98}
M. Cieplak, A.~Giacometti, A.~Maritan, A.~Rinaldo, I.~Rodriguez-Iturbe, and J.\,R.~Banavar,
\newblock Models of fractal river basins.
\newblock Journal of Statistical Physics, {\it 91}(1--2) 1--15 (1998).

\bibitem{PT00}
J.\,D.~Pelletier and D.\,L.~Turcotte, 
\newblock Shapes of river networks and leaves: are they statistically similar?
\newblock Philosophical Transactions of the Royal Society of London B: Biological Sciences,  {\it 355}(1394), 307--311 (2000).

\bibitem{LF07} 
B. Lashermes and E. Foufoula-Georgiou, 
\newblock Area and width functions of river networks: New results on multifractal properties.
\newblock Water Resour. Res., 43, W09405, (2007).

\bibitem{GW90}
V.\,K.~Gupta and E.\,C.~Waymire, 
\newblock Multiscaling properties of spatial rainfall and river flow distributions.
\newblock Journal of Geophysical Research: Atmospheres, 95(D3), 1999--2009 (1990).

\bibitem{GW}
H.\,W. Watson, F. Galton,  
\newblock On the probability of the extinction of families. 
\newblock The Journal of the Anthropological Institute of Great Britain and Ireland, 4, 138--144 (1875).

\bibitem{A01}
D.\,J. Aldous,  
\newblock Stochastic models and descriptive statistics for phylogenetic trees, from Yule to today. 
\newblock Statistical Science, 16(1), 23--34 (2001).

\bibitem{O92}
P.~Ossadnik,   
\newblock Branch order and ramification analysis of large diffusion-limited-aggregation clusters,
\newblock Physical Review A, {\it 45}(2), 1058 (1992).

\bibitem{DNS86}
G. Daccord, J. Nittmann, and H.\,E. Stanley, 
\newblock Radial viscous fingers and diffusion-limited aggregation: Fractal dimension and growth sites. 
\newblock Physical Review Letters, 56(4), 336 (1986).

\bibitem{FGD10}
E. Foufoula-Georgiou, V. Ganti, and W.\,E. Dietrich, 
\newblock A non-local theory for sediment transport on hillslopes. 
\newblock J. Geophys. Res. - Earth Surface, 115, F00A16 (2010).

\bibitem{SRF15}
A. Singh, L. Reinhardt, and E. Foufoula-Georgiou, 
\newblock Landscape reorganization under changing climatic forcing: results from an experimental landscape. 
\newblock Water Resour. Res., 51(6), 4320--4337 (2015).

\bibitem{WMPGC14}	
S.\,D. Willett, S.\,W. McCoy, J.\,T. Perron, L. Goren, and C. Y. Chen,  
\newblock Dynamic reorganization of river basins. 
\newblock Science, 343(6175) (2014).

\bibitem{MGM06}	
R. Mantilla, V. K. Gupta, and O. J. Mesa,  
\newblock Role of coupled flow dynamics and real network structures on Hortonian scaling of peak flows. 
\newblock Journal of Hydrology, 322(1-4), 155--167 (2006).

\bibitem{TLEZGRF17}
A. Tejedor, A. Longjas, D.\,A. Edmonds, I. Zaliapin, T.\,T. Georgiou, A. Rinaldo, and E. Foufoula-Georgiou, 
\newblock Entropy and optimality in river deltas. 
\newblock Proceedings of the National Academy of Sciences, 114(44), 11651--11656 (2017).

\bibitem{TLZF15}
A. Tejedor, A. Longjas, I. Zaliapin, and E. Foufoula-Georgiou, 
\newblock Delta channel networks: 1. A graph-theoretic approach for studying connectivity and steady state transport on deltaic surfaces.
\newblock Water Resources Research, 51(6), 3998--4018 (2015).

\bibitem{Ahlfors}
L.\,V.~Ahlfors, 
\newblock Complex analysis: an introduction to the theory of analytic functions of one complex variable. 
\newblock New York, London, 177 (1953).

\bibitem{BW07}
\newblock R. N.~Bhattacharya and E. C.~Waymire,
\newblock A basic course in probability theory (Vol. 69), Springer, New York  (2007).

\bibitem{Wilf}
H.\,S.~Wilf,  
\newblock Generatingfunctionology. 
\newblock Philadelphia, PA, USA (1992).

\end{thebibliography}

\end{document}